\documentclass[ reprint,  amsmath,amssymb,  aps, ]{revtex4-1}%
\usepackage[dvipdfmx]{graphicx,color}
\usepackage{dcolumn}
\usepackage{bm}
\usepackage{amsmath}
\usepackage{amsfonts}
\usepackage{amssymb}
\usepackage{mathrsfs}
\usepackage{relsize}
\usepackage{caption}
\providecommand{\U}[1]{\protect\rule{.1in}{.1in}}

\begin{document}

\preprint{APS/123-QED}

\title{Theory of Thermal Conductivity on Excitonic Insulator}

%\thanks{A footnote to the article title}

\author{Shinta Takarada}

%\altaffiliation[Also at ]{Physics Department, XYZ University.}

\author{Masao Ogata}
\author{Hiroyasu Matsuura}

%\email{Second.Author@institution.edu}%

\affiliation{
Department of Physics, University of Tokyo, Bunkyo-ku, 
Tokyo 113-0033, Japan}

%\collaboration{MUSO Collaboration}%

%\noaffiliation

\date{\today}

\begin{abstract}
We study the thermal conductivity in the excitonic insulator using a simple quasi one-dimensional two-band model consisting of electron and hole bands with the Coulomb interactions between these bands.
Based on the linear response theory within a mean-field scheme, we develop a method to identify the contributions to thermal conductivity driven by excitonic insulator.
It is found that there is an additional heat current operator owing to the excitonic phase transition, 
and that it gives contributions to thermal conductivity which are not expressed in the form of Sommerfeld-Bethe relations written in the form of an imaginary-time derivative of the electric current operator. 
Finally, we discuss the relationship between the newly-found additional contribution and the heat current carried by excitons.

\end{abstract}

\maketitle

\section{\label{sec:introduction}introduction}

An excitonic insulator (EI) is one of the interesting correlated phases 
of narrow-gap semiconductors and semimetals. 
It was proposed in the 60s that 
Coulomb interaction between conduction band electrons and valence band holes 
can lead to a spontaneous formation of excitons, and their condensation 
induces a nonconducting state~\cite{mott,blatt,knox,kohn,jerome,halperin}.
Various theoretical studies have been carried out on EI, 
and one of the most important and interesting aspects of these studies is that 
the excitonic theory can be transformed to the BCS theory 
by a particle-hole transformation~\cite{kohn,jerome,halperin}.
Then, various properties such as anisotropic band structure~\cite{zittartz4}, 
impurity effect~\cite{zittartz1}, transport properties~\cite{jerome,zittartz2,zittartz3}, 
and effect of magnetic field~\cite{fenton} have been extensively discussed 
in terms of similarity or symmetry with the superconducting theory.
However, no materials have been identified as EI at that time.

Recently, growing number of promising candidate materials for EI have actually been proposed, 
raising researchers' interest to study the EI phase. 
For example, ${\rm Tm}({\rm Se,Te})$~\cite{bucher,wachter2}, 
$1T$-${\rm TiSe}_2$~\cite{wilson, cercellier}, 
and ${\rm Ta}_2{\rm NiSe}_5$~\cite{tanise1,wakisaka,tanise2} 
have been proposed on the basis of various transport measurements~\cite{neuenschwander,wachter},   
angle-resolved photoemission spectroscopy (ARPES)~\cite{cercellier,tanise1,wakisaka}, 
and systematic elemental substitutions~\cite{zerogap}. 
In particular, after ${\rm Ta}_2{\rm NiSe}_5$ was proposed, 
many experiments have been performed on ${\rm Ta}_2{\rm NiSe}_5$, 
such as spectroscopic ellipsometry~\cite{ellipsometry}, 
observation of electron-phonon coupling and exciton-phonon coupling~\cite{epc1,epc2,epc3}, 
transport studies on bulk and thin-film samples~\cite{thin}, 
and study of electrical tuning of the EI ground state~\cite{tuning}.
The EI phase is also being actively studied theoretically, 
including calculation of the BCS-BEC crossover~\cite{bronold,crossover}, 
electron-phonon coupling~\cite{epc4}, 
spin-orbit coupling~\cite{soc}, 
and the topological EI states~\cite{topological}. 
With that, the verification of the theories and experiments has become increasingly important. 
It is interesting to note that some theoretical calculations 
reproduce the ARPES results~\cite{tanise2}, 
the superconductivity in the vicinity of the excitonic phase~\cite{tyoudendou,tyoudendouriron} and 
peculiar temperature dependence of orbital susceptibility~\cite{taijiritu,matsuura}.

The proposals of new candidate materials have also been pursued, 
such as semiconductor materials~\cite{handoutai,inas}, 
electron-hole bilayers~\cite{bilayer}, 
graphene~\cite{graphen1,graphen2} 
and iron-based superconductors~\cite{fe,fe2}. 
However, in actual materials, the EI phase often coexists 
with the charge density waves and staggered orbital orders~\cite{difficulty1,difficulty2}, 
and it is still difficult to determine the EI state by experiments.
Considering the verification of experimental and theoretical studies  
and the creation of new materials, 
it is important to develop methods other than photoemission spectroscopy 
to identify EI.
Thermal conductivity in EI is also one of the interesting topics. 
This is because the excitons do not have electronic charges, but have energies contributing to the heat current.
Indeed, it was observed that the thermal conductivity of ${\rm TmSe}_{0.45}{\rm Te}_{0.55}$, 
which is a candidate material of EI, 
shows unusual temperature dependence~\cite{wachter}.

When the thermal conductivity is written in the form of
an imaginary-time derivative of the electric current, 
as shown by Jonson and Mahan~\cite{jonson},
the following relations hold:
\begin{subequations}
\label{jonson}
\begin{align}
\label{jonson1}
L_{11}=&\int^{\infty}_{-\infty}d\epsilon
\left(-\frac{d\,f\left(\epsilon\right)}{d\epsilon}\right)
\sigma\left(\epsilon\right)
,\\
\label{jonson2}
L_{21}=&\int^{\infty}_{-\infty}d\epsilon
\,\frac{\epsilon-\mu}{e}
\left(-\frac{d\,f\left(\epsilon\right)}{d\epsilon}\right)
\sigma\left(\epsilon\right)
,\\
\label{jonson3}
L_{22}=&\int^{\infty}_{-\infty}d\epsilon
\left(\frac{\epsilon-\mu}{e}\right)^2
\left(-\frac{d\,f\left(\epsilon\right)}{d\epsilon}\right)
\sigma\left(\epsilon\right)
,\end{align}
\end{subequations}
where the linear response coefficients,
$L_{ij}\ \left(i,j=1,2\right)$,
are defined by~\cite{linear}
\begin{align}
\begin{split}
\bm{j}&= L_{11}\bm{E}+L_{12}\left(-\frac{\nabla T}{T}\right)
,\\
\bm{j}^Q&=L_{21}\bm{E}+L_{22}\left(-\frac{\nabla T}{T}\right)
.\end{split}
\end{align}
Here, $\bm{j}$, $\bm{j}^Q$, $\bm{E}$, $T$, and $\nabla T$
are electric current density, heat current density,
electric field, temperature, and temperature gradient, respectively, 
and $e$ is electron charge ($e<0$) and
$f\left(\epsilon\right)=
1/\left(e^{\beta\left(\epsilon-\mu\right)}+1\right)$
is the Fermi distribution function, where $\beta=1/k_BT$, and $\mu$ and $k_B$ are the chemical potential and Boltzmann constant.
Due to Onsager's reciprocal theorem, $L_{12}=L_{21}$ holds.
The electrical conductivity is $L_{11}$,
and the thermal conductivity $\kappa$,
which is defined as the ratio of $\bm{j}_Q$ to $-\nabla T$
under the $\bm{j}=0$ condition, is given by
\begin{align}
\label{kappa}
\kappa=\left(L_{22}-L_{21}L_{12}/L_{11}\right)/T.
\end{align}

Since excitons do not carry electrical charge, $L_{11}$ in Eq.~(\ref{jonson1}) 
should be due to quasiparticles and not due to excitons.
Correspondingly, $L_{21}$ and $L_{22}$ in Eqs.~(\ref{jonson2}) and (\ref{jonson3}) 
should be also due to quasiparticles.
Therefore, if the excitons, when they are condensed below transition temperature, 
contribute to heat current, then there must be additional contributions to $L_{21}$ and $L_{22}$ 
that are not expressed as in Eqs.~(\ref{jonson2}) and (\ref{jonson3}).
To clarify the discussion in the following, 
we call Eqs.~(\ref{jonson2}) and (\ref{jonson3}) as Sommerfeld-Bethe (SB) relation for $L_{21}$~\cite{fukuyama}
and SB relation for $L_{22}$, because Eqs.~(\ref{jonson2}) and (\ref{jonson3}) can be obtained
from the Boltzmann's equation~\cite{sb}.
In the previous studies~\cite{zittartz3, kurihara}, the thermal conductivity was discussed on the basis of Eqs.~(\ref{jonson1}) $\sim$ (\ref{jonson3}).
Thus, the exciton contributions were not taken into account.
%Since excitons do not contribute to the electrical conductivity, 
%$\sigma\left(\epsilon\right)$ in Eq.~(\ref{jonson1}) is 
%due to the quasiparticles in EI. 
%Therefore, the thermal conductivity obtained from Eq.~(\ref{jonson3}) is also due to the quasiparticles. 
%However, in principle, excitons can contribute to the thermal conductivity 
%since they carry energies. 
%This effect is not included in the previous studies~\cite{zittartz3, kurihara}. 

In this paper, we study the thermal conductivity in a model for EI to identify the additional contribution derived from EI, 
which is beyond the SB relations.
First, we microscopically obtain the heat current operator in a model for EI~\cite{ohta,matsuura}, or a simple quasi one-dimensional two-band model.
Then, we study the linear response theory~\cite{kubo,luttinger} of 
$L_{11}$, $L_{21}$, and $L_{22}$ 
within a mean-field scheme to introduce the order parameters of EI.
We will show that, by considering two types of nearest-neighbor interactions,
additional heat current operators owing to the EI phase transition
exist and give the contributions in $L_{21}$ and $L_{22}$ 
which are not expressed in the form of Eqs.~(\ref{jonson2}) and (\ref{jonson3}). 
%Since these contributions are related to the order parameters of EI, 
%we identify them as contributions of excitons to heat current. 
Finally, we discuss the relationship between this additional contribution and the heat current due to excitons.

Theoretically, the validity of the SB relations, Eqs.~(\ref{jonson1}) $\sim$ (\ref{jonson3}), 
has been discussed~\cite{jonson,kontani,fukuyama}.
Jonson and Mahan showed that, in the presence of potential (including random potentials) 
and the electron-phonon interaction, the SB relations holds, 
except for a single term in the heat current operator which is due to the electron-phonon interaction~\cite{jonson}.
Later, Kontani showed that the presence of the Hubbard interaction does not break the SB relations using the Jonson-Mahan’s method~\cite{kontani}.
Recently it is shown that the presence of the finite range Coulomb interaction breaks 
the SB relations~\cite{fukuyama}.
The present paper is an extension of~\cite{fukuyama} 
to the case where a phase transition occurs 
in the presence of finite range interactions.

%In the EI phase, the thermal conductivity increases with decreasing temperature.
%This temperature dependence is different from that expected in the BCS theory 
%in which the thermal conductivity decreases as the temperature decreases 
%from the transition temperature\cite{bcs,zittartz3}.
 
This paper is organized as follows. 
In Sect.~2, we introduce a model Hamiltonian to study the thermal conductivity in EI, 
and then develop a method for calculating the electronic state 
based on the mean-field approximation.
In Sect.~3, we derive heat current operators on this model microscopically. 
In Sect.~4, we clarify the contribution of the order parameters of EI to the thermal conductivity, 
and in Sect.~5, we discuss the validity of SB relations on this model, 
and the relationship between the additional contribution and the heat cuurent due to excitons.
Finally, Sect.~6 is devoted for the conclusion.

\section{\label{sec:model}Two-Band Model and Electronic State of Excitonic Insulator based on Mean-field Approximation}

\begin{figure}[t]
  \begin{center}
   \includegraphics[width=7cm]{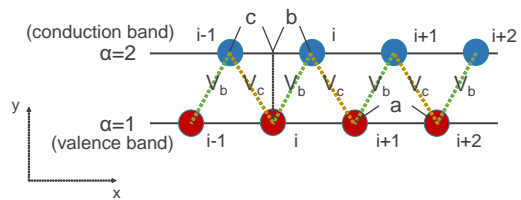}
   \captionsetup{justification=raggedright}
   \caption{Schematic picture of two-band model with Coulomb interactions between chains. 
Red and blue circles indicate the lattice sites, $\alpha$ is the index of the chain, 
$b$ is the difference between the $x$-coordinates of the $i$-th site of two chains, 
and $c$ is the difference between $i$-th and ($i$-1)-th sites. 
$V_b$ and $V_c$ are the Coulomb interactions between the two chains.
}
 \label{fig:2band}
  \end{center}
\end{figure}

To study EI, in this paper, 
we use a following two-band model Hamiltonian as shown in Fig.~\ref{fig:2band}~\cite{matsuura}:
\begin{align}
\hat{\mathscr{H}} =& \sum_i [
\,( t \,\hat{c}^{\dagger}_{1,i}\,\hat{c}_{1,i+1} 
- t'\,\hat{c}^{\dagger}_{2,i}\,\hat{c}_{2,i+1} +{\rm h.c.} \,)\,
\nonumber\\
&+ \left( -\epsilon - \mu \right)\hat{n}_{1,i}
+\left( \epsilon - \mu \right) \hat{n}_{2,i}
+V_b \, \hat{n}_{1,i}\, \hat{n}_{2,i} 
\nonumber\\
&+V_c \, \hat{n}_{1,i}\, \hat{n}_{2,i-1} 
\,]\,
+\sum_{i,j,\alpha}
V_{\rm{imp}}\left( x_{\alpha,i} - X_j \right) \hat{n}_{\alpha,i}
,\label{H}
\end{align}
where $t$ ($t'$) and $\epsilon$ are a transfer integral between 
the nearest-neighbor sites in the chain $\alpha = 1$ ($\alpha =2$), 
and a one-body level of $i$-th site, respectively. 
$\hat{c}_{\alpha,i}$ ($\hat{c}_{\alpha,i}^\dagger$) is an annihilation (creation) operator 
at the $i$-th site of chain $\alpha$ where the spin degrees of freedom are neglected.
$V_b$ and $V_c$ indicate the Coulomb interaction 
between $i$-th site in the chain $\alpha =1$ and $i$-th site in the chain $\alpha =2$ 
and that between $i$-th site in the chain $\alpha = 1$ 
and ($i$-1)-th site in the chain $\alpha = 2$. 
Figure \ref{fig:2band} shows the positions of sites.
$b$ ($c$) is the difference between the $x$-coordinates of the $i$-th site 
in the chain $\alpha = 1$ and the $i$-th (($i$-1)-th) site in the chain $\alpha = 2$.
We set the lattice constant $a$ as $a=b+c$.
$V_{\rm imp}\left(x_{\alpha,i}-X_j\right)$ is the randomly distributed impurity potential where $x_{\alpha,i}$ and
$X_j$ are the $x$-coordinates of the position of the site and randomly distributed impurities, respectively. 

The similar effective model has been suggested in Ref.~\cite{matsuura}. 
The model of Ref.~\cite{matsuura} corresponds to that for $c=a$, $b=0$ and $V_c=0$ in Eq.~(\ref{H}).
As discussed below, the finite $b$ and $V_c$ are important 
to obtain the additional heat current due to the Coulomb interaction.  
%\begin{align}
%\hat{n}_{1,i}\,\hat{n}_{2,j} \rightarrow &
%- \langle \hat{c}^{\dagger}_{2,j}\,\hat{c}_{1,i} \rangle
%\hat{c}^{\dagger}_{1,i}\,\hat{c}_{2,j}
%- \langle \hat{c}^{\dagger}_{1,i}\,\hat{c}_{2,j} \rangle
%\hat{c}^{\dagger}_{2,j}\,\hat{c}_{1,i}
%\nonumber\\
%&+ \langle \hat{c}^{\dagger}_{2,j}\,\hat{c}_{1,i} \rangle
%\langle \hat{c}^{\dagger}_{1,i}\,\hat{c}_{2,j} \rangle
%,\label{heikinba}
%\end{align}

Using a mean-field approximation for EI 
and assuming that the order parameters are independent of the sites, i.e., 
\begin{align}
\begin{split}
\label{delta}
\Delta _b =& -\frac{V_b}{N}
\sum_{i}\langle \hat{c}^{\dagger}_{2,i}\,\hat{c}_{1,i} \rangle =-\frac{V_b}{N} \sum_k
\langle \hat{c}^{\dagger}_{2,k}\,\hat{c}_{1,k} \rangle e^{-ikb}
,\\
\Delta _c =&  -\frac{V_c}{N}
\sum_{i}\langle \hat{c}^{\dagger}_{2,i-1}\,\hat{c}_{1,i} \rangle =-\frac{V_c}{N} \sum_k
\langle \hat{c}^{\dagger}_{2,k}\,\hat{c}_{1,k} \rangle e^{ikc}
,\end{split}
\end{align}
with $N$ being the total number of sites, 
we obtain the following mean field Hamiltonian of EI 
\begin{align}
\hat{\mathscr{H}}_{\rm MF} =&
\sum_k \{
\left( 2t\,{\rm cos}ka -\epsilon -\mu \right)
\hat{c}^{\dagger}_{1,k}\,\hat{c}_{1,k}
\nonumber\\
&+ \left( -2t'\,{\rm cos}ka + \epsilon - \mu \right)
\hat{c}^{\dagger}_{2,k}\,\hat{c}_{2,k}
\nonumber\\
&+ (\,\Delta_k\,
\hat{c}^{\dagger}_{1,k}\,\hat{c}_{2,k}
+ {\rm h.c.})\}
\nonumber\\
&+ \sum_{k,q,\alpha} \frac{1}{N} \,\rho_{\rm imp} \left(q\right)
v_{\rm imp}\left(q\right)
\hat{c}^{\dagger}_{\alpha,k+q}\,\hat{c}_{\alpha,k}
,\label{Hmfk}
\end{align}
where
$\Delta_k \equiv\, \Delta_b \,e^{ikb} + \Delta_c \,e^{-ikc}$, 
$\rho_{\rm imp}\left(q\right)\equiv \sum_ie^{-iqX_i}$
and $v_{\rm imp}\left(q\right)\equiv \sum_i
e^{-iq\left(x_{\alpha,i}-X\right)}\,V_{\rm imp}\left(x_{\alpha,i}-X\right)$.

By diagonalizing this Hamiltonian except for the impurity potential, 
we obtain  
\begin{align}
\hat{\mathscr{H}}_{\rm eff} = 
\sum_k \left( E_{k+}\,\hat{p}^{\dagger}_{k+}\,\hat{p}_{k+} 
+  E_{k-}\,\hat{p}^{\dagger}_{k-}\,\hat{p}_{k-} \right)
,\label{Heff}
\end{align}
where $\hat{p}_{k\pm}$ is an annihilation operator of a quasiparticle, 
and the quasiparticle energy is given by 
\begin{align}
E_{k\pm}=& \left(t-t'\right)\,{\rm cos}ka - \mu 
\nonumber\\
&\pm \sqrt{\left\{ \left(t+t'\right)\,{\rm cos}ka - \epsilon \right\}^2 
+ \left|\Delta_k\right|^2}
.\label{Ek}
\end{align}
The unitary matrix satisfying 
$\left( \begin{array}{c}
\hat{c}_{1,k} \\ \hat{c}_{2,k}
\end{array} \right) \equiv
\,\bm{U}\left( \begin{array}{c}
\hat{p}_{k+} \\ \hat{p}_{k-}
\end{array} \right) $ is 
\begin{align}
\bm{U}=
\frac{1}{\sqrt{2X_k}}
\left( \begin{array}{cc}
\frac{\Delta_k}{\sqrt{X_k-Y_k}} 
& \frac{\Delta_k}{\sqrt{X_k+Y_k}}  
\\
\frac{X_k-Y_k}{\sqrt{X_k-Y_k}} 
& \frac{-X_k-Y_k}{\sqrt{X_k+Y_k}}
\end{array} \right),
\label{U}
\end{align}
where $X_k$ and $Y_k$ are 
\begin{align}
\begin{split}
X_k &\equiv
\sqrt{\left\{ \left(t+t'\right)\,{\rm cos}ka - \epsilon \right\}^2 
+ \left|\Delta_k\right|^2}, 
\\
Y_k &\equiv \left(t+t'\right)\,{\rm cos}ka - \epsilon
.\end{split}
\end{align}

Using the effective Hamiltonian of Eq.~(\ref{Heff}), 
the self-consistent equations to obtain the order parameters of EI becomes 
\begin{align}
\Delta _b =& 
-\frac{V_b}{2N} \sum_k \left( \Delta_b + \Delta_c \,e^{-ika} \right) 
\frac{ f\left(E_{k+}\right) - f\left(E_{k-}\right) }{X_k}
,\nonumber\\
\Delta_c =&
-\frac{V_c}{2N} \sum_k \left( \Delta_b \,e^{ika} + \Delta_c \right) 
\frac{ f\left(E_{k+}\right) - f\left(E_{k-}\right)}{X_k}
.\label{gap}
\end{align}
%where $f\left(E\right)=\frac{1}{e^{\beta E}+1}$ 
%is the Fermi distribution function.
The chemical potential is now included in $E_{k\pm}$. 
Since we focus on the half-filling case in this paper, the chemical potential is determined by 
\begin{align}
\label{mu}
\frac{1}{N} \sum_k \left\{ f\left(E_{k+}\right) + f\left(E_{k-}\right) \right\}=1
.\end{align}

Solving Eqs.~(\ref{gap}) and (\ref{mu}) self-consistently, we can obtain $\Delta_b$\, ,\, $\Delta_c$ and $\mu$.
It is to be noted that the order parameters can be complex.
However, we find that the phase difference between $\Delta_b$ and $\Delta_c$ does not occur in the parameters we used in this paper. 
Thus, we set the order parameters as real numbers.

\begin{figure}[t]
 \begin{minipage}{0.49\hsize}
  \begin{center}
   \includegraphics[width=4.2cm]{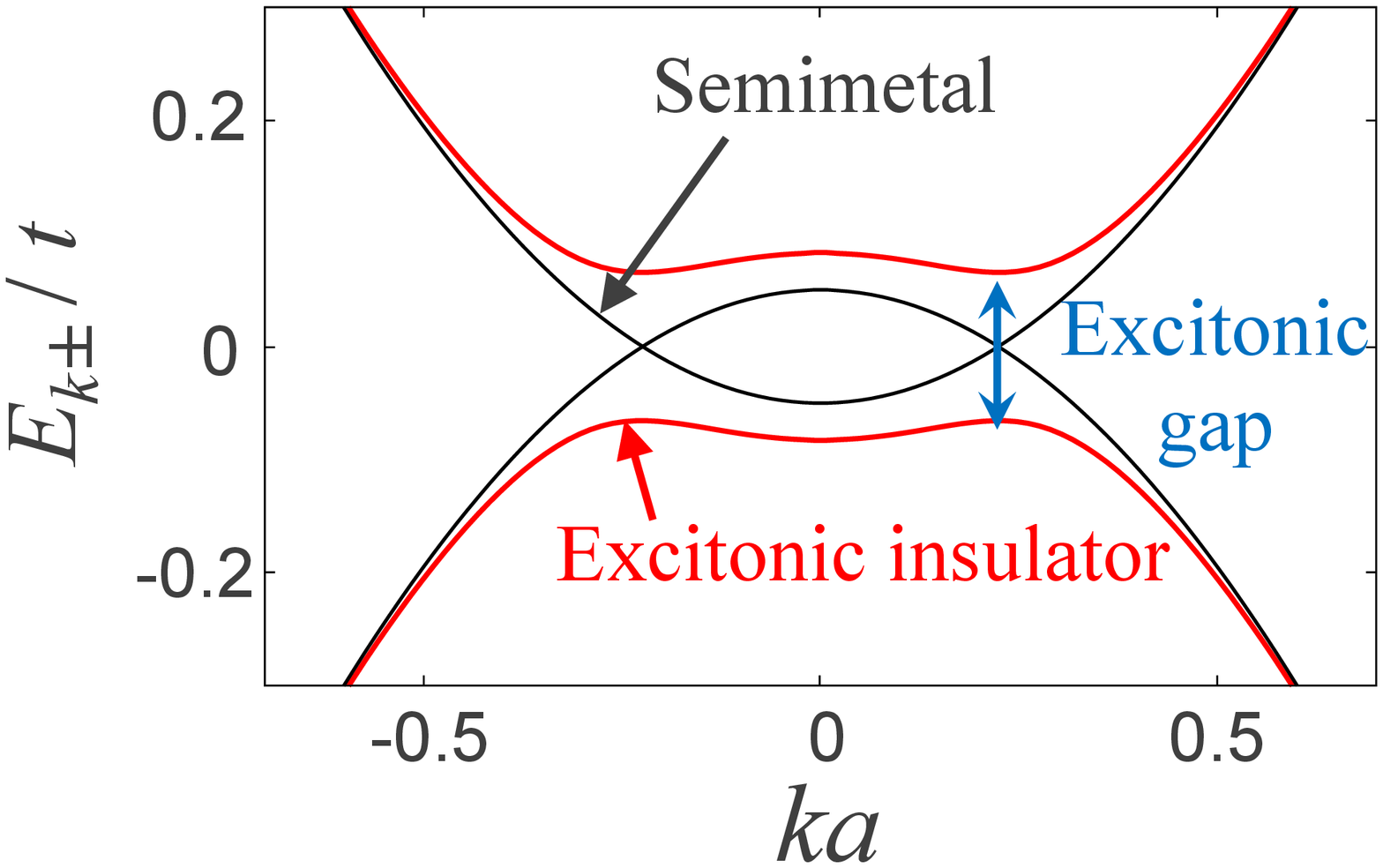}
   \captionsetup{labelformat=empty,labelsep=none}
   \caption*{(a)}
  \end{center}
 \end{minipage}
 \begin{minipage}{0.49\hsize}
  \begin{center}
   \includegraphics[width=4.2cm]{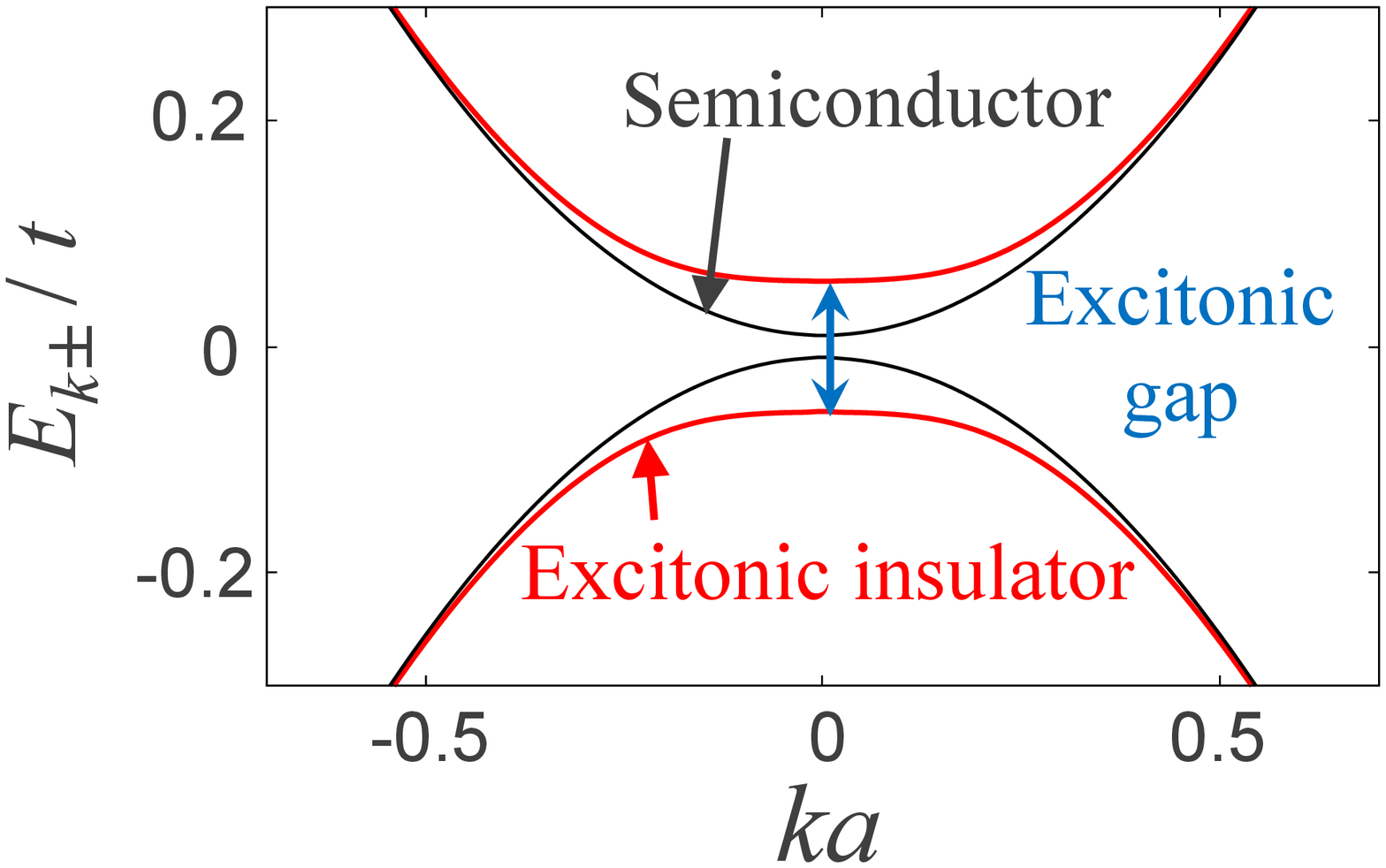}
   \captionsetup{labelformat=empty,labelsep=none}
   \caption*{(b)}
  \end{center}
 \end{minipage}
\\
 \begin{minipage}{0.49\hsize}
  \begin{center}
   \includegraphics[width=4.2cm]{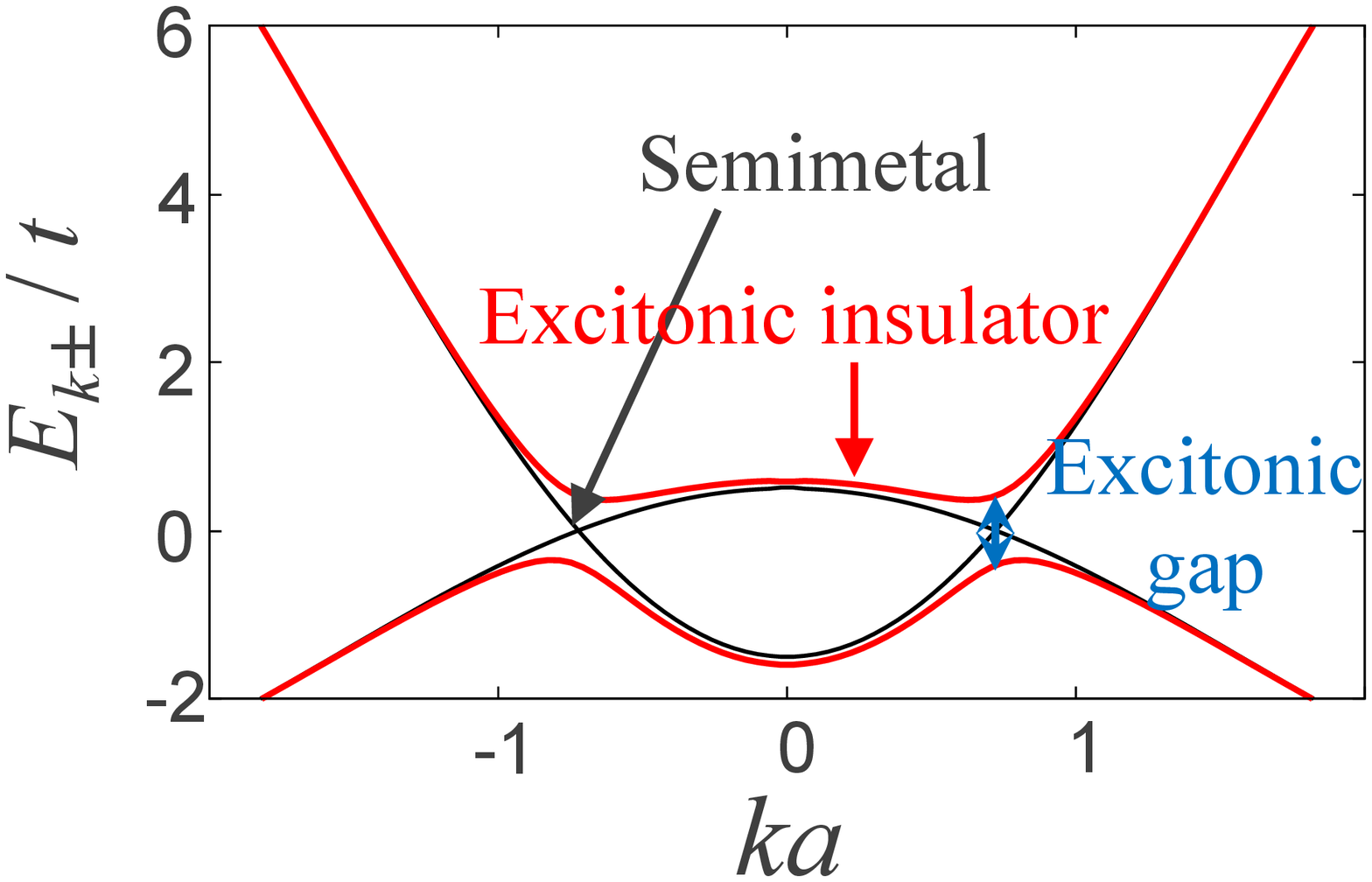}
   \captionsetup{labelformat=empty,labelsep=none}
   \caption*{(c)}
  \end{center}
 \end{minipage}
 \begin{minipage}{0.49\hsize}
  \begin{center}
   \includegraphics[width=4.2cm]{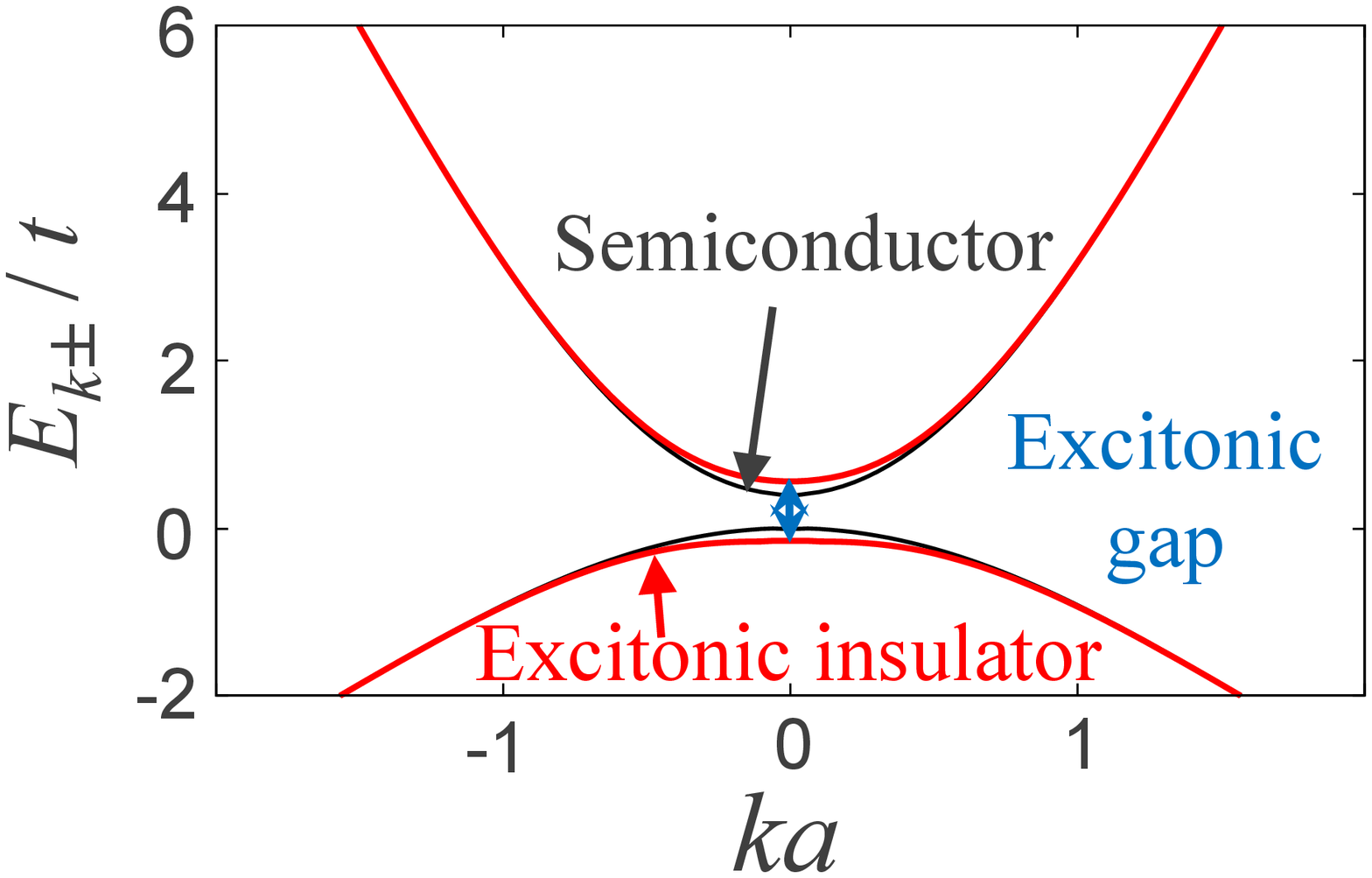}
   \captionsetup{labelformat=empty,labelsep=none}
   \caption*{(d)}
  \end{center}
 \end{minipage}
\captionsetup{justification=raggedright}
\caption{Dispersion relations of normal state (black line) and 
EI (red line) at $T/t = 0$.
The parameters are set as (a) $t'=t$, $V_b =  V_c = 0.5t$, and $\epsilon = 1.95t$, (b) $t'=t$, $V_b =  V_c = 0.5t$, and $\epsilon = 2.01t$, 
(c) $t'=3t$, $V_b =  V_c = 2t$ and $\epsilon = 3t$, and (d) $t'=3t$, $V_b =  V_c = 2t$ and $\epsilon = 4.2t$. 
}
\label{fig:bunsan}
\end{figure}

Figure \ref{fig:bunsan} shows the dispersion relations of normal state (black lines) and 
EI (red lines) at $T/t = 0$ for several values of Hamiltonian parameters. 
We set $t'=t$,  $V_b=V_c=0.5t$, and $\epsilon =1.95t$ for Fig.~\ref{fig:bunsan}(a) 
and $\epsilon = 2.01t$ for Fig.~\ref{fig:bunsan}(b), 
and $t'=3t$, $V_b=V_c=2t$, and $\epsilon =3t$ for Fig.~\ref{fig:bunsan}(c) 
and $\epsilon = 4.2t$ for Fig.~\ref{fig:bunsan}(d), respectively. 
The dispersion relations of the normal state indicate the semimetallic state for Figs.~\ref{fig:bunsan}(a) and (c), 
or the semiconducting state for Figs.~\ref{fig:bunsan}(b) and (d).
When we consider the order parameters of EI, all the electronic states become the semiconducting states. 

\begin{figure}[t]
 \begin{minipage}{0.49\hsize}
  \begin{center}
   \includegraphics[width=4.2cm]{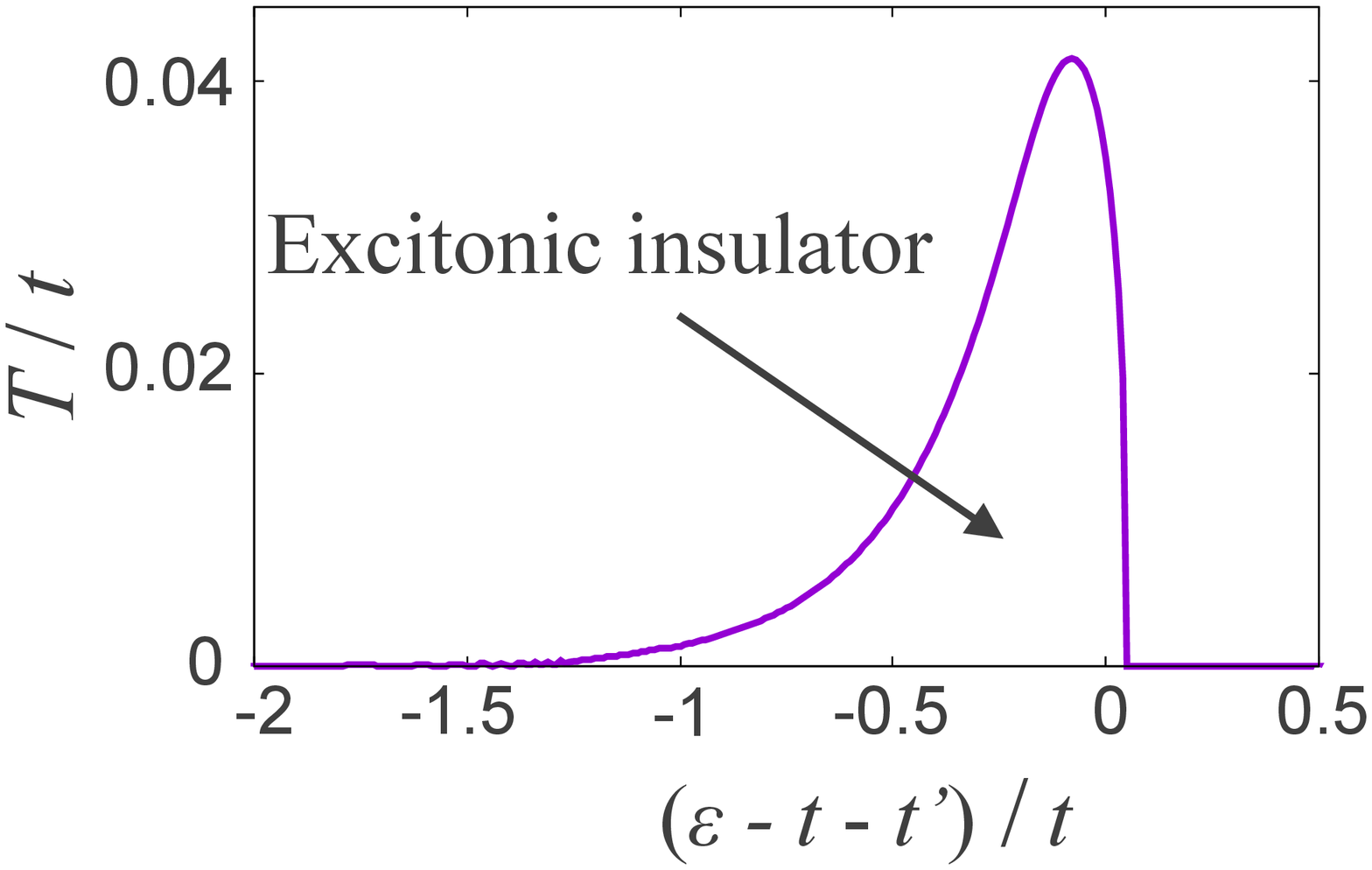}
   \captionsetup{labelformat=empty,labelsep=none}
   \caption*{(a)}
  \end{center}
 \end{minipage}
 \begin{minipage}{0.49\hsize}
  \begin{center}
   \includegraphics[width=4.2cm]{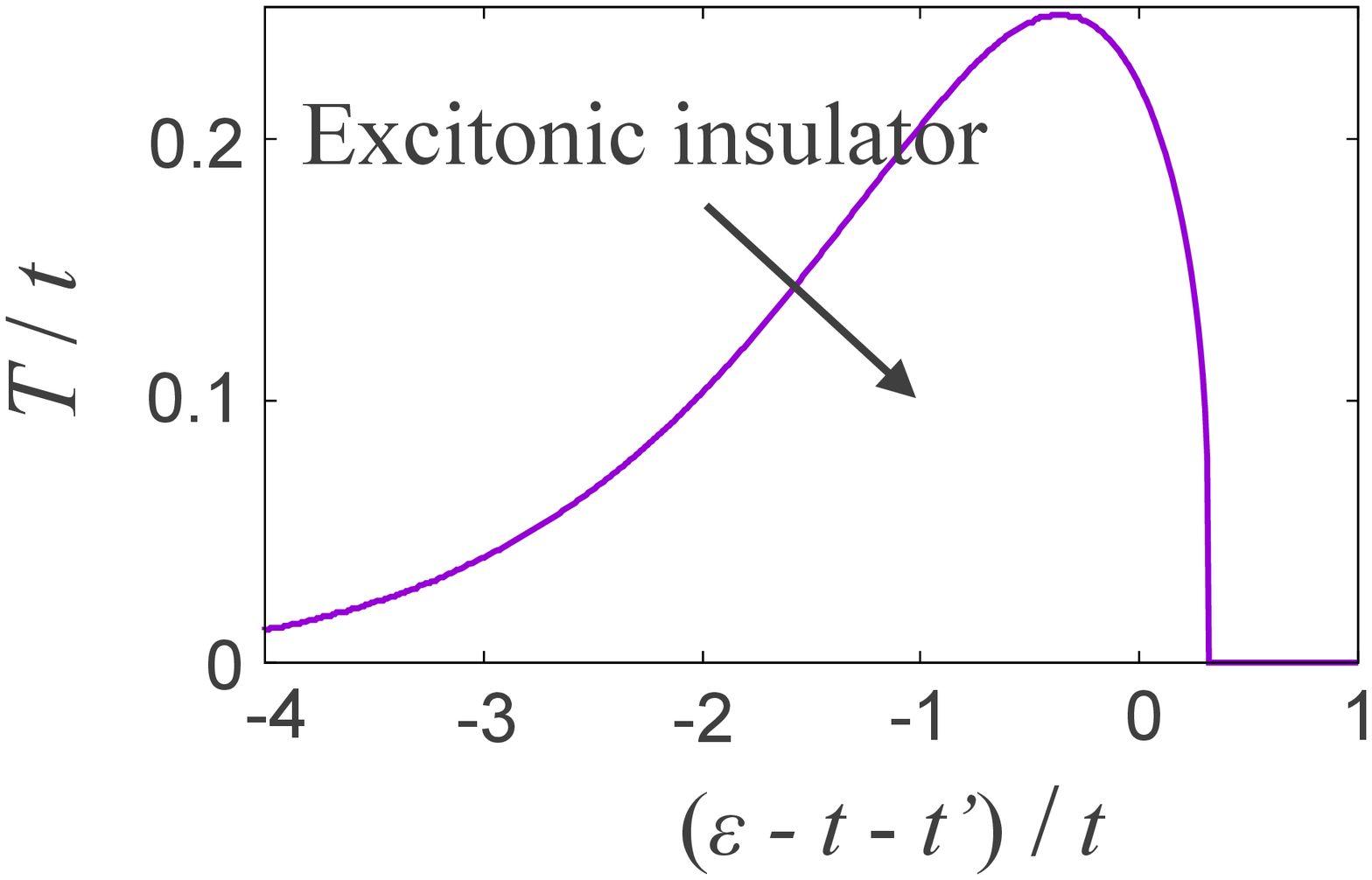}
   \captionsetup{labelformat=empty,labelsep=none}
   \caption*{(b)}
  \end{center}
 \end{minipage}
\captionsetup{justification=raggedright}
\caption{$\epsilon$ dependence of the transition temperature at (a) $t'= t$ and $V_b = V_c = 0.5t$ 
and (b) $t'=3t$ and $V_b = V_c = 2t$.
}
\label{fig:souzu}
\end{figure}

Figures \ref{fig:souzu}(a) and (b) show 
the $\epsilon$ dependence of the transition temperature of EI.
The parameters of  Fig.~\ref{fig:souzu}(a) (Fig.~\ref{fig:souzu}(b)) are the same as that of Fig.~\ref{fig:bunsan}(a) (Fig.~\ref{fig:bunsan}(c)).
As shown in Fig.~\ref{fig:souzu}(a) and (b), the transition temperature has a maximum at $(\epsilon -t-t^\prime)/t \sim 0$.
This $\epsilon$ behavior is the same as the result of Ref.~\cite{matsuura}.

\section{\label{sec:heat} Electric and heat current operators}

The electric current density operator $\hat{\bm{j}}\left(\bm{r}\right)$ 
and the heat current density operator $\hat{\bm{j}}^Q\left(\bm{r}\right)$ 
are derived from the continuity equations~\cite{linear}
\begin{align}
\frac{d\hat{\rho}\left(\bm{r}\right)}{dt}
+{\rm div}\hat{\bm{j}}\left(\bm{r}\right)=0
,\;\;
\frac{d\hat{h}\left(\bm{r}\right)}{dt}
+{\rm div}\hat{\bm{j}}^Q\left(\bm{r}\right)=0
,\label{con_eq}
\end{align}
where $\hat{\rho}\left(\bm{r}\right)$ and 
$\hat{h}\left(\bm{r}\right)$ are the charge density 
and the Hamiltonian density 
($\hat{\mathscr{H}}=\int d\bm{r}\,\hat{h}\left(\bm{r}\right)$), respectively.

In the preset model (Eq.~(\ref{H})), 
the total electric current operator, 
$\hat{J}=\int d\bm{r}\hat{j}\left(\bm{r}\right)$, 
is given by 
\begin{align}
\label{denryu}
\hat{J}=\sum_{k,\alpha}j_{\alpha}\left(k\right)\,
\hat{c}^{\dagger}_{\alpha,k}\, \hat{c}_{\alpha,k}
,\end{align}
with
\begin{align}
j_{1}\left(k\right)\equiv -\frac{2ea}{\hbar} t \,{\rm sin}ka
,\;\;
j_{2}\left(k\right)\equiv \,\frac{2ea}{\hbar} t'\, {\rm sin}ka
.\end{align}
On the other hand, the total heat current operator, 
\\$\hat{J}^Q=\int d\bm{r}\hat{j}^Q\left(\bm{r}\right)$,
becomes 
\begin{align}
\hat{J}^Q=&\sum_{k,\alpha}j^Q_{\alpha\alpha}\left(k\right)
\hat{c}^{\dagger}_{\alpha,k}\,\hat{c}_{\alpha ,k}
\nonumber\\
&+\sum_{k,k',q}j^Q_{V}\left(k,k',q\right)
\hat{c}^{\dagger}_{1,k+q}\,\hat{c}_{1 ,k}\,
\hat{c}^{\dagger}_{2,k'-q}\,\hat{c}_{2 ,k'}
\nonumber\\
&+\sum_{k,q,\alpha} j^Q_{{\rm imp}\alpha}\left(k,q\right)
\hat{c}^{\dagger}_{\alpha,k+q}\,\hat{c}_{\alpha,k},
\label{neturyuuenzansi}
\end{align}
with
\begin{align}
&j^Q_{11}\left(k\right)\equiv
-\frac{2a}{\hbar}t\,{\rm sin}ka
\left(2t\,{\rm cos}ka- \epsilon -\mu\right)
,\nonumber\\
&j^Q_{22}\left(k\right)\equiv
\frac{2a}{\hbar}t'\,{\rm sin}ka
\left(-2t'\,{\rm cos}ka + \epsilon -\mu\right)
,\nonumber\\
&j^Q_{V}\left(k,k',q\right)
\nonumber\\
\equiv&
-\frac{a}{\hbar}
\left(t-t'\right)\frac{1}{N}\Bigl\{\left( V_be^{iqb}+V_ce^{-iqc}\right)
{\rm cos}\frac{q}{2}a\,{\rm sin}(k+\frac{q}{2})a
\nonumber\\
&+\left( V_be^{iq(b+\frac{a}{2})}+V_ce^{-iq(c+\frac{a}{2})}\right)
{\rm sin}(k'-\frac{q}{2})a \Bigr\}
,\nonumber\\
&j^Q_{{\rm imp}1}\left(k,q\right) 
\nonumber\\
\equiv&-\frac{2a}{\hbar}t\frac{1}{N}
\,\rho_{\rm imp}\left(q\right)v_{\rm imp}\left(q\right)
{\rm cos}\frac{q}{2}a\,{\rm sin}(k+\frac{q}{2})a
,\nonumber\\
&j^Q_{{\rm imp}2}\left(k,q\right) 
\nonumber\\
\equiv&\frac{2a}{\hbar}t'\frac{1}{N}
\,\rho_{\rm imp}\left(q\right)v_{\rm imp}\left(q\right)
{\rm cos}\frac{q}{2}a\,{\rm sin}(k+\frac{q}{2})a 
.\label{jcoefficient}
\end{align}
The derivations of these operators together with the Hamiltonian density 
are shown in Appendix A.
While $\hat{J}$ in Eq.~(\ref{denryu}) is a one-body current operator, 
the heat current operator $\hat{J}^Q$ in Eq.~(\ref{neturyuuenzansi}) is the many-body operator 
in the presence of the Coulomb interaction~\cite{fukuyama}.

As discussed in Appendix B, if we start from the mean field Hamiltonian of Eq.~(\ref{Hmfk}), 
we obtain a different expression of heat current operator.
However, the heat current operator in Eq.~(\ref{neturyuuenzansi}) 
should be used to study the thermal conductivity, because it satisfies the continuity equation (\ref{con_eq})
without any approximations.

\section{\label{conductivity}
Additional heat current contribution in the EI phase} 

In this section, we study thermal conductivity based on the linear response theory 
within the mean-field approximation to clarify the additional contributions in the EI phase, 
which do not satisfy the SB relation in Eqs.~(\ref{jonson}).

In the mean-field approximation (\ref{Hmfk}), 
the one-body Green's function is given as   
\begin{align}
\mathscr{G}^0_{\alpha \alpha '}\left( k,\tau\right)&\equiv 
-\left\langle T_{\tau}\left[\, \hat{p}_{k\alpha}\left(\tau\right) 
\hat{p}^{\dagger}_{k\alpha '}\left(0\right) \right] \right\rangle
,\\
\bm{G}^{(0)}\left( k,i\epsilon_n \right)&\equiv
\left(\begin{array}{cc}
\mathscr{G}^0_{\scriptscriptstyle ++}\left( k,i\epsilon_n \right) 
&\mathscr{G}^0_{\scriptscriptstyle +-}\left( k,i\epsilon_n \right) 
\\
\mathscr{G}^0_{\scriptscriptstyle -+}\left( k,i\epsilon_n \right)
&\mathscr{G}^0_{\scriptscriptstyle --}\left( k,i\epsilon_n \right)
\end{array}\right)
\nonumber\\
&=\left(\begin{array}{cc}
\frac{1}{i\epsilon_n - E_{k+}}& 0 \\
0& \frac{1}{i\epsilon_n - E_{k-}}
\end{array}\right)
,\end{align}
where $T_{\tau}$ is the standard $\tau$-ordering operator, 
and $\epsilon_n =(2n +1)\pi k_BT$ where $n$ is an integer.
Taking into account the self-energy due to the impurity scattering 
\begin{align}
\label{sigmakeisankekka}
\bm{\Sigma}\left(k,i\epsilon_n\right)
=&-i\,{\rm sign}\left(\epsilon_n\right)
\left( \begin{array}{cc}
\Gamma_1&\Gamma_3\\
\Gamma_3&\Gamma_2
\end{array}\right)
,\end{align}
where $\Gamma_1$, $\Gamma_2$ and $\Gamma_3$ are real independent of $\epsilon_n$, 
Dyson's equation 
%for the full Green's function is 
%\begin{align}
%\label{dyson}
%\bm{G}\left( k,i\epsilon_n\right) =
%\bm{G}^0 \left( k,i\epsilon_n\right)+
%\bm{G}^0 \left( k,i\epsilon_n\right)
%\bm{\Sigma} \left( k,i\epsilon_n\right)
%\bm{G} \left( k,i\epsilon_n\right)
%.\end{align}
%Then, the full green's function is obtained as
leads to
\begin{widetext}
\begin{align}
\bm{G}\left(k,i\epsilon_n\right)
={\scriptstyle \frac{1}
{\left( i\epsilon_n - E_{k+}+i\,{\rm sign}\left(\epsilon_n\right)\Gamma_1\right)
\left( i\epsilon_n - E_{k-}+i\,{\rm sign}\left(\epsilon_n\right)\Gamma_2\right)
+\Gamma_3^2}}
\left( \begin{array}{cc}
i\epsilon_n - E_{k-}+i\,{\rm sign}\left(\epsilon_n\right)\Gamma_2
& -i\,{\rm sign}\left(\epsilon_n\right)\Gamma_3
\\
-i\,{\rm sign}\left(\epsilon_n\right)\Gamma_3
& i\epsilon_n - E_{k+}+i\,{\rm sign}\left(\omega_n\right)\Gamma_1
\end{array}\right).
\label{G}
\end{align}
\end{widetext}
In this paper, the values of $\Gamma_1$ $\Gamma_2$ and $\Gamma_3$ are
estimated by calculating the self-energy in the absence of interactions
from Dyson's Equation as follows
\begin{align}
\begin{split}
\Gamma_1=&\frac{an_iv^2}{2X_k}
\left(\frac{X_k+Y_k}{\sqrt{4t^2-\left(\epsilon+\mu\right)^2}}
+\frac{X_k-Y_k}{\sqrt{4t'^2-\left(\epsilon-\mu\right)^2}}\right)
,\\
\Gamma_2=&\frac{an_iv^2}{2X_k}
\left(\frac{X_k-Y_k}{\sqrt{4t^2-\left(\epsilon+\mu\right)^2}}
+\frac{X_k+Y_k}{\sqrt{4t'^2-\left(\epsilon-\mu\right)^2}}\right)
,\\
\Gamma_3=&\frac{an_iv^2}{2X_k}
\left(\frac{\left|\Delta_k\right|}{\sqrt{4t^2-\left(\epsilon+\mu\right)^2}}
+\frac{\left|\Delta_k\right|}{\sqrt{4t'^2-\left(\epsilon-\mu\right)^2}}\right)
.\end{split}
\end{align}
Here, $n_i$ is the impurity density,
and we assumed that $v_{\rm imp}\left(q\right)=v$, ignoring the $q$ dependence of
$v_{\rm imp}\left(q\right)$.

We also apply the mean-field approximation to the heat current operator to obtain 
\begin{align}
\hat{J}^Q_{{\rm MF}}=&
\sum_{k,\alpha ,\alpha '}
j^Q_{\alpha\alpha '}\left(k\right)
\hat{c}^{\dagger}_{\alpha,k}\,\hat{c}_{\alpha ' ,k}
\nonumber\\
&+\sum_{k,q,\alpha} j^Q_{{\rm imp}\alpha}\left(k,q\right)
\hat{c}^{\dagger}_{\alpha,k+q}\,\hat{c}_{\alpha,k},
\label{neturyuuenzansiheikinba}
\end{align}
where $j^Q_{12}$ and $j^Q_{21}$ are
\begin{align}
&j^Q_{12}\left(k\right)=
\overline{j^Q_{21}\left(k\right)}
\nonumber\\
\equiv&
-\frac{a}{\hbar}
\left(t-t'\right)\Bigl\{\Delta_k\,{\rm sin}ka
+\frac{i}{4}\left(\frac{V_c}{V_b}\Delta_b\,e^{-ikc}
-\frac{V_b}{V_c}\Delta_c\,e^{ikb}
\right)
\nonumber\\
&-\frac{i}{2}\left(\Delta_b \,e^{ikb}-\Delta_c \,e^{-ikc}
\right){\rm cos}ka
\Bigr\}.
\label{j12}
\end{align}
We will proceed with the calculation assuming that the effect of impurities 
is sufficiently small.
From Eq.~(\ref{jcoefficient}), 
it can be seen that the heat current 
$j^Q_{{\rm imp}1}\left(k,q\right)$ 
and $j^Q_{{\rm imp}2}\left(k,q\right)$
due to impurities only have effects of the order of 
$\mathcal{O}\left(v_{\rm imp}\right)$ 
or $\mathcal{O}\left(v_{\rm imp}^2\right)$,
which is smaller than the other heat current.
Therefore, the heat current due to impurities 
will be ignored in following calculations. 
We also ignore the vertex correction due to impurities as well, 
since it only affects the current and heat current operators of the order of 
$\mathcal{O}\left(v_{\rm imp}^2\right)$.

Using the annihilation and creation operators of quasiparticle, 
$\hat{p}^{\dagger}_{k\alpha}$ and $\hat{p}_{k\alpha}$, 
the electric current and heat current operators are written as 
\begin{align}
\begin{split}
\hat{J}=&\sum_k \left(\begin{array}{cc}
\hat{p}^{\dagger}_{k+}& \hat{p}^{\dagger}_{k-}
\end{array}\right)\bm{\Gamma}\left(k\right)
\left(\begin{array}{c}
\hat{p}_{k+}\\ \hat{p}_{k-}
\end{array}\right)
,\\
\hat{J}^Q_{\rm MF}
=&\sum_k \left(\begin{array}{cc}
\hat{p}^{\dagger}_{k+}& \hat{p}^{\dagger}_{k-}
\end{array}\right)\bm{\Gamma}^Q\left(k\right)
\left(\begin{array}{c}
\hat{p}_{k+}\\ \hat{p}_{k-}
\end{array}\right)
,\end{split}
\end{align}
where  
$\bm{\Gamma}=\bm{U}^{\dagger}
\left(\begin{array}{cc}j_{1}&0\\0&j_{2}\end{array}\right)\bm{U}$, 
$\bm{\Gamma}^Q=\bm{U}^{\dagger}
\left(\begin{array}{cc}j^Q_{11}&j^Q_{12}
\\j^Q_{21}&j^Q_{22}\end{array}\right)\bm{U}$. 

$L_{11}$ is obtained by the $\hat{J}$-$\hat{J}$ correlation 
\begin{align}
&\Phi_{11}\left(\tau\right)
=\frac{1}{aN}\left\langle T_{\tau}
\left[\,\hat{J}\left(\tau\right)
\hat{J}\left(0\right)\right]\right\rangle
\nonumber\\
\label{soukan5syou}
=&-\frac{1}{aN}\sum_{k}
{\rm Tr}\left[\bm{\Gamma}\left(k\right)\bm{G}\left(k,\tau\right)
\bm{\Gamma}\left(k\right)\bm{G}\left(k,-\tau\right)\right]
,\end{align}
and its Fourier transform
\begin{align}
\label{soukan5syoufourier}
\Phi_{11}\left(i\omega_{\lambda}\right)
=&-\frac{k_BT}{aN}\sum_{k}\sum_{n}
{\rm Tr}[\bm{\Gamma}\left(k\right)\bm{G}\left(k,i\epsilon_n\right)
\nonumber\\
&\times\bm{\Gamma}\left(k\right)
\bm{G}\left(k,i\epsilon_n-i\omega_{\lambda}\right)]
.\end{align}
Here, $\omega_{\lambda}=2\lambda\pi k_BT$ and $\lambda$ is an integer.

\begin{figure}[t]
 \begin{minipage}{0.44\hsize}
  \begin{center}
   \includegraphics[width=3.5cm]{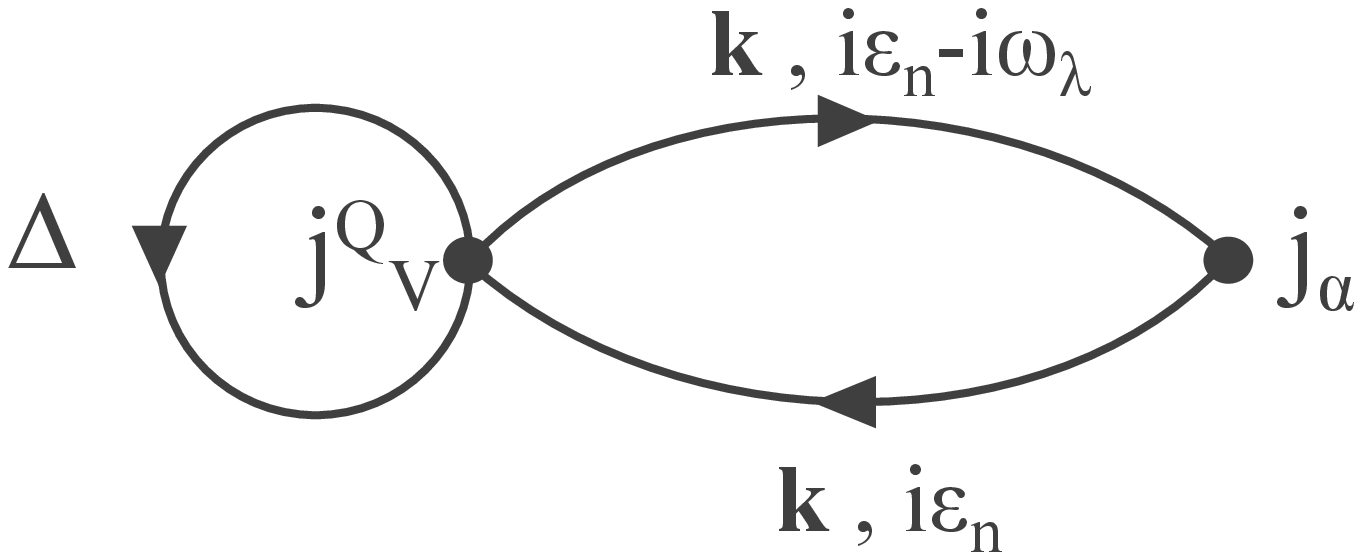}
   \captionsetup{labelformat=empty,labelsep=none}
   \caption*{(a)}
  \end{center}
 \end{minipage}
 \begin{minipage}{0.54\hsize}
  \begin{center}
   \includegraphics[width=4cm]{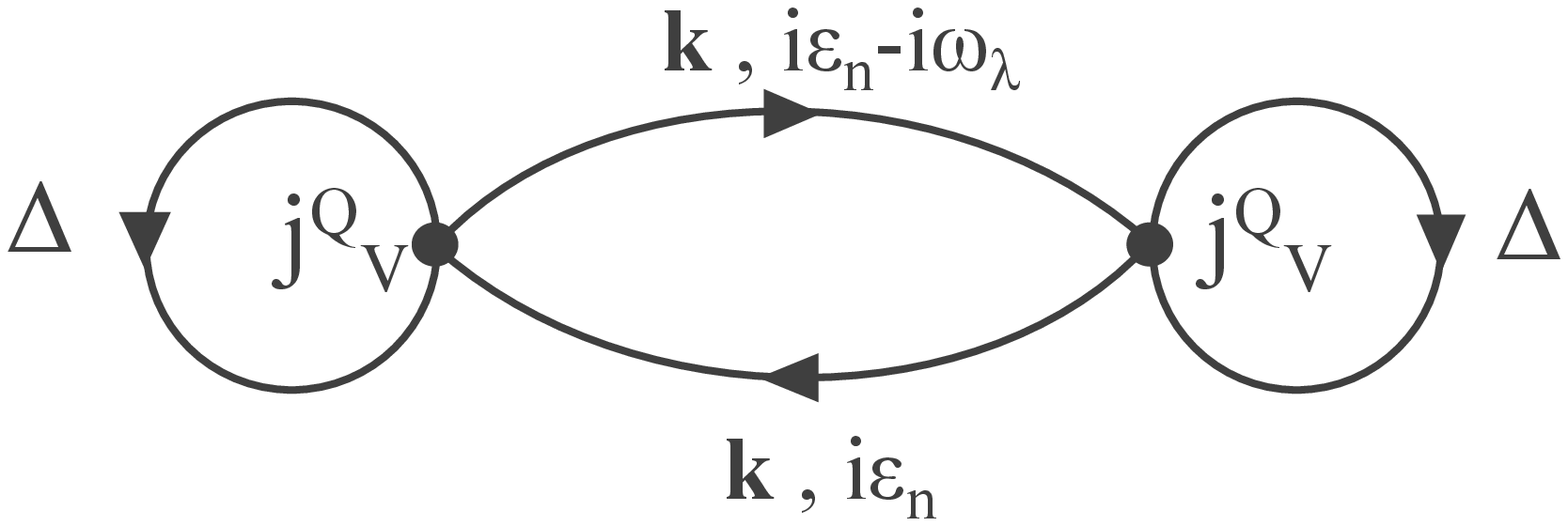}
   \captionsetup{labelformat=empty,labelsep=none}
   \caption*{(b)}
  \end{center}
 \end{minipage}
\captionsetup{justification=raggedright}
\caption{Feynman diagrams for heat currents including interactions.}
\label{fig:diagram}
\end{figure}

By performing the analytic continuation 
($i\omega_{\lambda}\rightarrow \hbar\omega+i\delta$), 
we can obtain 
\begin{align}
L_{11}=\lim_{\omega \to 0}
\frac{\Phi_{11}\left(\hbar\omega +i\delta\right)-\Phi_{11}\left(0\right)}
{i\omega}
.\label{l11keisanhou}
\end{align}
Similarly, $L_{21}$, $L_{22}$ can be calculated by 
\begin{align}
\Phi_{21}\left(i\omega_{\lambda}\right)
=&-\frac{k_BT}{aN}\sum_{k}\sum_{n}
{\rm Tr}[\bm{\Gamma}^Q\left(k\right)\bm{G}\left(k,i\epsilon_n\right)
\nonumber\\
&\times\bm{\Gamma}\left(k\right)
\bm{G}\left(k,i\epsilon_n-i\omega_{\lambda}\right)]
,\label{l21keisanhou}\\
\Phi_{22}\left(i\omega_{\lambda}\right)
=&-\frac{k_BT}{aN}\sum_{k}\sum_{n}
{\rm Tr}[\bm{\Gamma}^Q\left(k\right)\bm{G}\left(k,i\epsilon_n\right)
\nonumber\\
&\times\bm{\Gamma}^Q\left(k\right)
\bm{G}\left(k,i\epsilon_n-i\omega_{\lambda}\right)].
\label{l22keisanhou}\end{align}
Equations.~(\ref{l21keisanhou}) and (\ref{l22keisanhou}) correspond to the Feynman diagrams shown in Fig. \ref{fig:diagram}.

We can see that the dominant contributions are in the order of 
$\mathcal{O}\left(1/\Gamma_1\right)$,  $\mathcal{O}\left(1/\Gamma_2\right)$
and $\mathcal{O}\left(1/\Gamma_3\right)$. 
Then, ignoring the higher order of 
$\mathcal{O}\left(\Gamma_1\right)$,  $\mathcal{O}\left(\Gamma_2\right)$ 
and $\mathcal{O}\left(\Gamma_3\right)$, 
we obtain the correlation functions as follows.  
\begin{subequations}\label{l}
\begin{align}
&L_{11}
\nonumber\\
=&\frac{e^2a^2}{4\pi\hbar}\int dk
\left(-f'\left(E_{k+}\right) 
\frac{\sigma_{k+}^2}{\Gamma_1}
-f'\left(E_{k-}\right)
\frac{\sigma_{k-}^2}{\Gamma_2}\right)
,\\
&L_{21}
\nonumber\\
=&\frac{ea^2}{4\pi\hbar}\int dk\left(
-E_{k+}f'\left(E_{k+}\right)
\frac{\sigma_{k+}^2}{\Gamma_1}
-E_{k-}f'\left(E_{k-}\right)
\frac{\sigma_{k-}^2}{\Gamma_2}\right)
\nonumber\\
&+\frac{ea^2}{4\pi\hbar}\int dk\left(
-f'\left(E_{k+}\right)
\frac{g_k\sigma_{k+}}{\Gamma_1}
+f'\left(E_{k-}\right)
\frac{g_k\sigma_{k-}}{\Gamma_2}\right)
,\\
&L_{22}
\nonumber\\
=&\frac{a^2}{4\pi\hbar}\int dk\left(
-E_{k+}^2f'\left(E_{k+}\right)
\frac{\sigma_{k+}^2}{\Gamma_1}
-E_{k-}^2f'\left(E_{k-}\right)
\frac{\sigma_{k-}^2}{\Gamma_2}\right)
\nonumber\\
&+\frac{a^2}{2\pi\hbar}\int dk\Bigl(
-E_{k+}f'\left(E_{k+}\right)
\frac{g_k\sigma_{k+}}{\Gamma_1}
\nonumber\\
&+E_{k-}f'\left(E_{k-}\right)
\frac{g_k\sigma_{k-}}{\Gamma_2}\Bigr)
\nonumber\\
&+\frac{a^2}{4\pi\hbar}\int dk\left(
-f'\left(E_{k+}\right)
\frac{g_k^2}{\Gamma_1}
-f'\left(E_{k-}\right)
\frac{g_k^2}{\Gamma_2}\right)
,\end{align}
\end{subequations}
where $\sigma_{k\pm}$ and $g_k$ are
\begin{eqnarray}
\sigma_{k\pm} &\equiv& \frac{{\rm sin}ka}{X_k}
\left\{\left(X_k\pm Y_k\right)t
-\left(X_k\mp Y_k\right)t'\right\}, \\
g_k &\equiv&
\frac{t-t'}{4X_k}\Bigl\{\left( \frac{V_c}{V_b} \Delta_b^2
+\frac{V_b}{V_c}\Delta_c^2\right) {\rm sin}ka
\nonumber\\
&&+2\Delta_b\Delta_c\,{\rm sin}2ka\Bigr\}
.\end{eqnarray}
Derivations of Eqs.~(\ref{l}) are given in Appendix C.

$L_{21}$ and $L_{22}$ differ from $L_{11}$ in that 
$g_k$ appears in the integrand. 
In other words, the first term in $L_{21}$ and $L_{22}$ are expressed by the SB relations. 
In contrast, the second term in $L_{21}$ and 
the second and third terms in $L_{22}$ are not expressed by the SB relations.
Therefore, these terms give the additional contributions to the thermal conductivity in the EI phase.
This is the main result of this paper, 
and we will discuss it more in detail in the next section.

\begin{figure}[t]
 \begin{minipage}{0.49\hsize}
  \begin{center}
   \includegraphics[width=4.2cm]{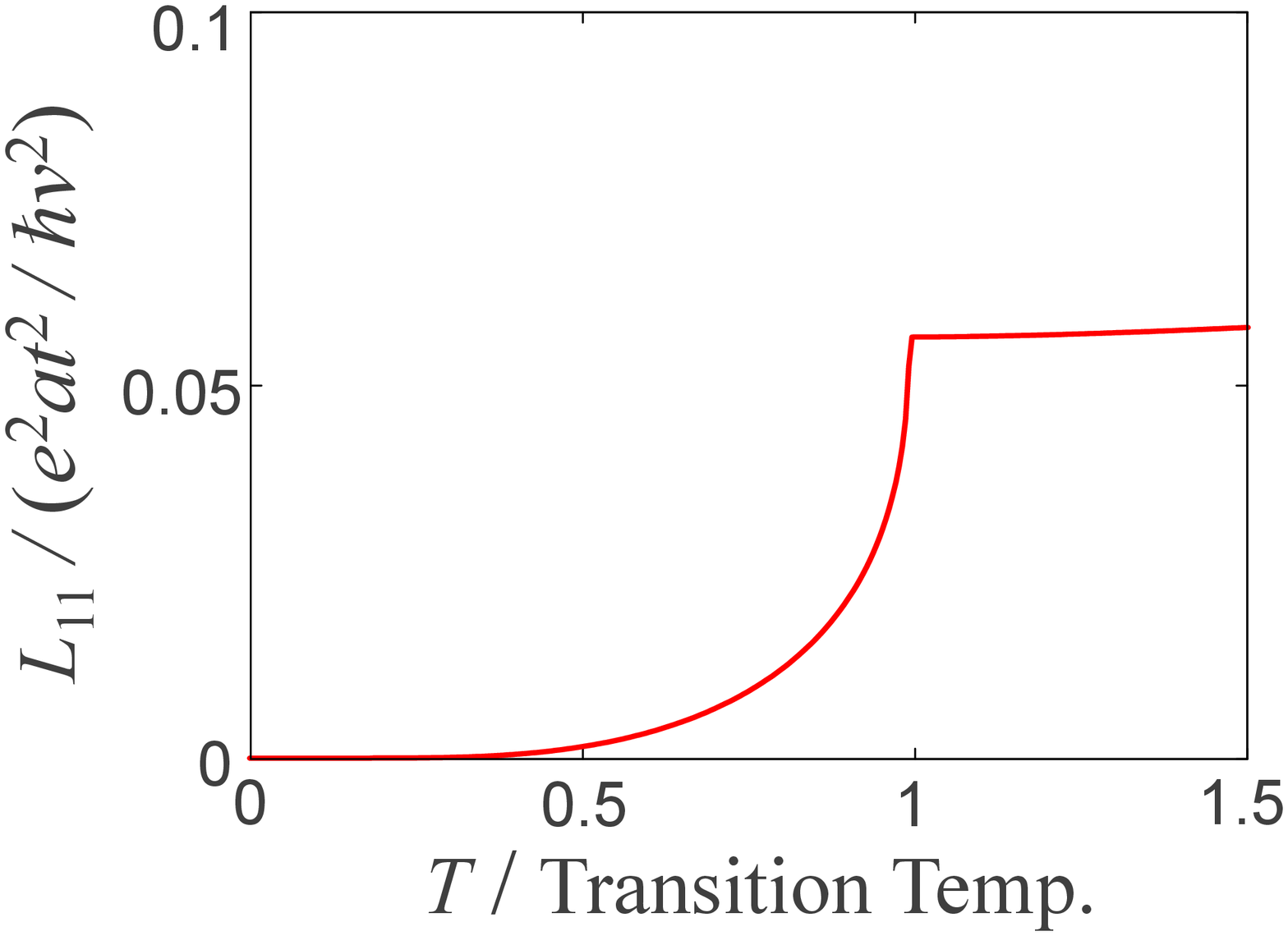}
   \captionsetup{labelformat=empty,labelsep=none}
   \caption*{(a)}
  \end{center}
 \end{minipage}
 \begin{minipage}{0.49\hsize}
  \begin{center}
   \includegraphics[width=4.2cm]{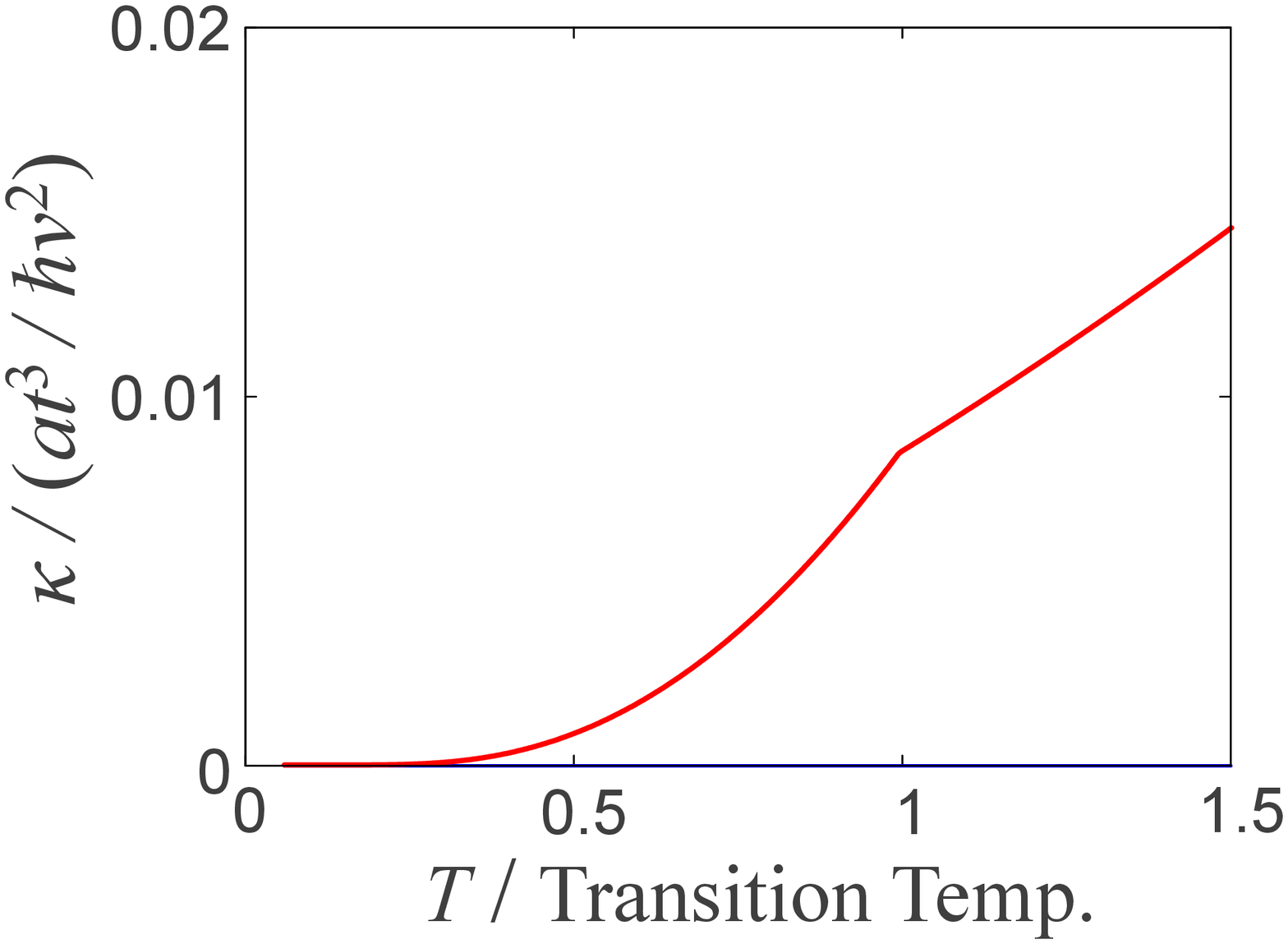}
   \captionsetup{labelformat=empty,labelsep=none}
   \caption*{(b)}
  \end{center}
 \end{minipage}
\\
 \begin{minipage}{0.49\hsize}
  \begin{center}
   \includegraphics[width=4.2cm]{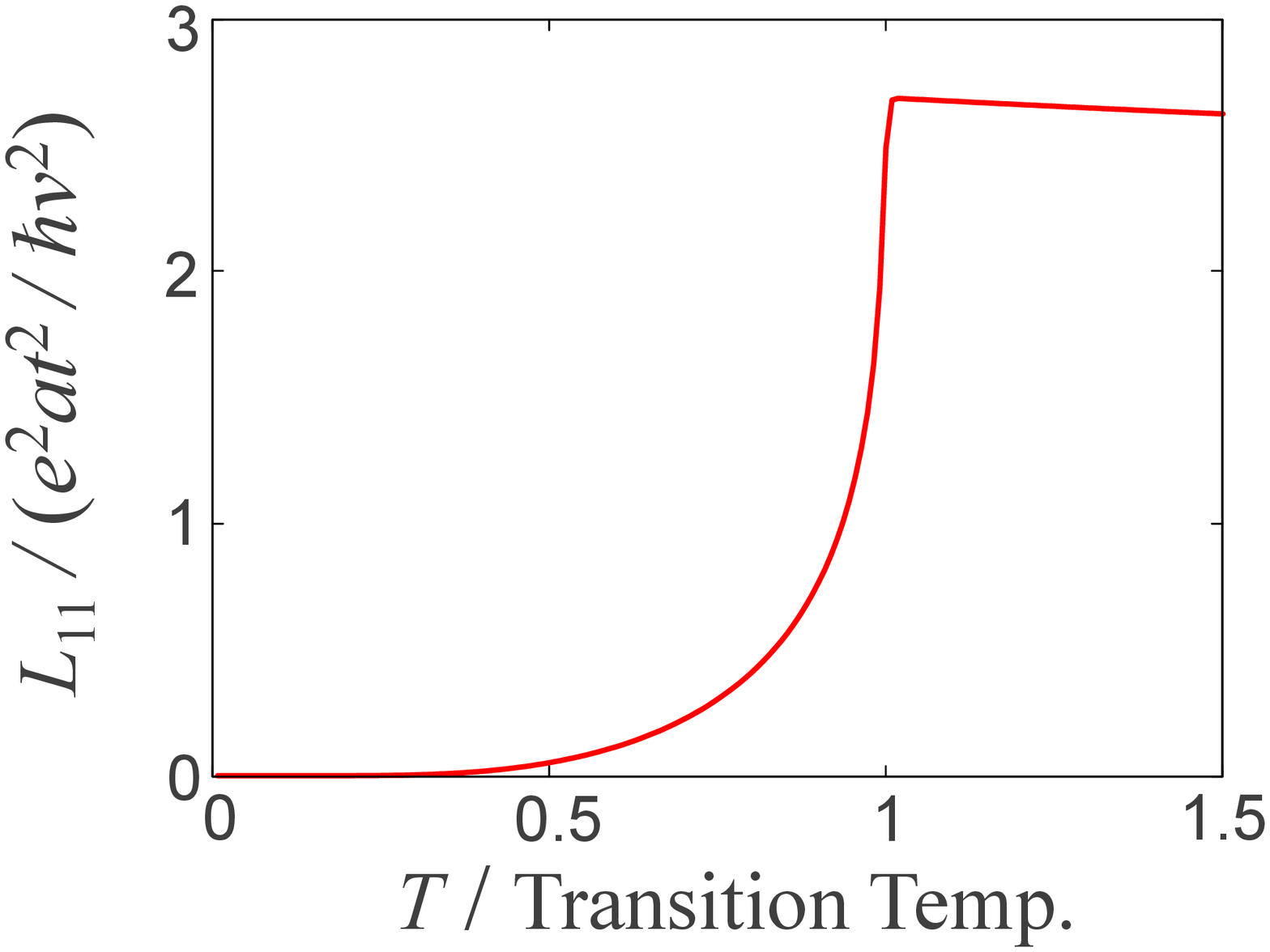}
   \captionsetup{labelformat=empty,labelsep=none}
   \caption*{(c)}
  \end{center}
 \end{minipage}
 \begin{minipage}{0.49\hsize}
  \begin{center}
   \includegraphics[width=4.2cm]{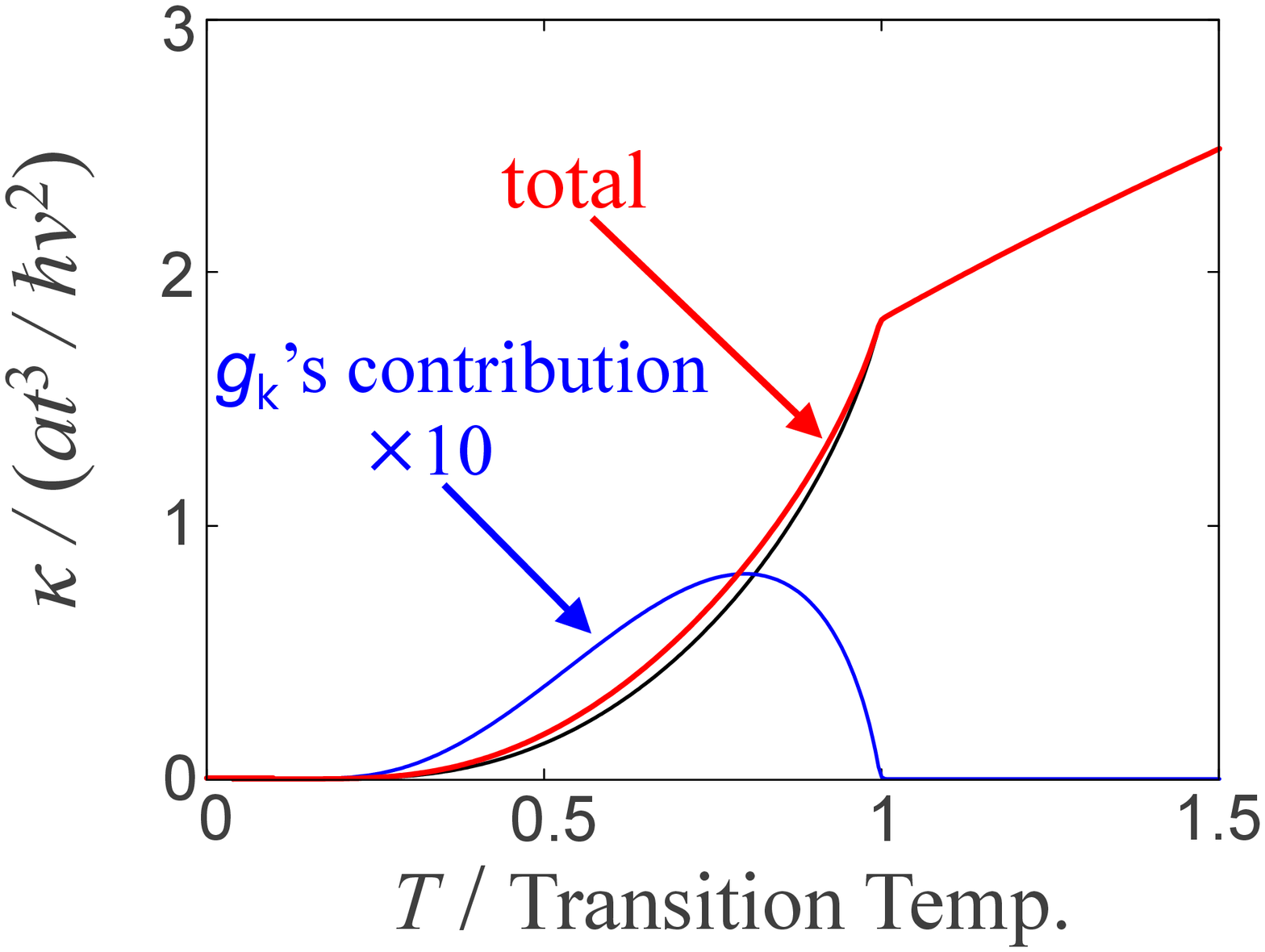}
   \captionsetup{labelformat=empty,labelsep=none}
   \caption*{(d)}
  \end{center}
 \end{minipage}
\captionsetup{justification=raggedright}
\caption{Temperature dependence on (a) $L_{11}$ and 
(b) $\kappa$
at $t' = t$, $V_b = V_c = 0.5t$, $\epsilon = 1.95t$, and 
(c) $L_{11}$ and (d) $\kappa$ at 
$t' =3t$, $V_b = V_c = 2t$, $\epsilon = 3t$. 
Blue, black, and red line indicates the contribution from
$g_k\times10$, other terms, and total, respectively.
}\label{fig:kekka}
\end{figure}

Before going into details, let us see the magnitude of this newly-found additional contributions.
Figures \ref{fig:kekka}(a) - (b) (\ref{fig:kekka}(c) - (d)) show the temperature dependence of 
$L_{11}$ and $\kappa$ at $t'=t$, $V_b=V_c=0.5t$, and $\epsilon =1.95t$ 
(at $t'=3t$, $V_b=V_c=2t$, and $\epsilon =3t$).
The portion of $\kappa$ that is related to 
$g_k$ is indicated by the blue line, 
the others by the black line, and the total by the red line.

It is found that the contribution of $g_k$ to the thermal conductivity is present 
below the transition temperature. 
When $t=t'$ the thermal conductivity follows the SB relation because $g_k=0$ 
(Fig.~\ref{fig:kekka}(b)), 
but when $t\neq t'$, it may be visible depending on the parameters 
(Fig.~\ref{fig:kekka}(d)).
This is the thermal conductivity produced by 
the additional heat current driven by EI.

\section{Discussions}

As mentioned in Sect.~1, when the heat current operator, 
$\widetilde{\hat{J}^Q}\left(\tau\right)$,
can be written in the form of an imaginary-time derivative of
the electric current
\\ ($\hat{J}\left(\tau,\tau'\right)=
\sum_{k}
\left\{j\left(k\right)
\hat{c}^{\dagger}_{k}\left(\tau\right)\hat{c}_{k}\left(\tau'\right)\right\}$) as
\begin{align}
\widetilde{\hat{J}^Q}\left(\tau\right) =\lim_{\tau' \to \tau}\frac{1}{2}\left(
\frac{\partial}{\partial \tau}-\frac{\partial}{\partial \tau'}
\right) \hat{J}\left(\tau,\tau'\right),
\end{align}
$L_{11}$, $L_{21}$, and $L_{22}$ are expressed by Eqs.~(\ref{jonson})~\cite{jonson,fukuyama},
where $\hat{c}_{k}\left(\tau\right)$ is an annihilation operator
with an imaginary-time dependence.

In the present case of mean-field approximation, 
if we consider that the imaginary-time derivative of $\tau$ and $\tau '$ 
are given by the mean field Hamiltonian (\ref{Hmfk}) as 
\begin{align}
\begin{split}
\frac{\partial}{\partial \tau}\hat{J}\left(\tau,\tau'\right)
=\sum_k j\left( k\right)\left[\hat{\mathscr{H}_{\rm MF}}\left(\tau\right),
\hat{c}^{\dagger}_k\left(\tau\right)\right]
\hat{c}_k\left(\tau'\right), \\
\frac{\partial}{\partial \tau'}\hat{J}\left(\tau,\tau'\right)
=\sum_k j\left( k\right)\hat{c}^{\dagger}_k\left(\tau\right)
\left[\hat{\mathscr{H}_{\rm MF}}\left(\tau'\right),\hat{c}_k\left(\tau'\right)\right].
\end{split}
\end{align}
We obtain 
\begin{align}
\widetilde{\hat{J}^Q}\left(\tau\right)=&\lim_{\tau' \to \tau}\frac{1}{2}\left(
\frac{\partial}{\partial \tau}-\frac{\partial}{\partial \tau'}
\right) \hat{J}\left(\tau,\tau'\right)
\nonumber\\
=&\sum_{k,\alpha ,\alpha '}
j'^Q_{\alpha\alpha '}\left(k\right)
\hat{c}^{\dagger}_{\alpha,k}\left(\tau\right)
\hat{c}_{\alpha ',k}\left(\tau\right)
\nonumber\\
&+\sum_{k,q,\alpha}
j^Q_{{\rm imp}\alpha}\left(k,q\right)
\hat{c}^{\dagger}_{\alpha,k+q}\left(\tau\right)
\hat{c}_{\alpha,k}\left(\tau\right)
,\end{align}
where
\begin{align}
j'^Q_{11}\left(k\right)\equiv& j^Q_{11}\left(k\right)
,\;\;
j'^Q_{22}\left(k\right)\equiv j^Q_{22}\left(k\right)
,\nonumber\\
j'^Q_{12}\left(k\right)
=&\overline{j'^Q_{21}\left(k\right)}
\equiv-\frac{a}{\hbar}\left(t-t'\right)\Delta_k \, {\rm sin}ka
.\label{jikanbibun}
\end{align}
When we use $\widetilde{\hat{J}^Q}$ instead of $\hat{J}^Q$, 
we obtain the first term in $L_{21}$ and $L_{22}$ in Eqs.~(\ref{l}). 
Therefore, this is consistent with the argument by Jonson-Mahan~\cite{jonson} 
that the heat current operator $\widetilde{\hat{J}^Q}$ leads to SB relations.

In other words, the contributions which violate the SB relations appears
in $L_{21}$ and $L_{22}$, that is $g_k$, 
from the second and third terms in $j^Q_{12}$ in Eq.~(\ref{jcoefficient}), 
which originate from the two types of Coulomb interactions.
Furthermore, this $g_k$ is represented by the order parameters of EI, 
and it contributes not to the electrical conductivity
but to the thermal conductivity below the transition temperature.
%Therefore, this will represent the heat transport by the excitons.

Although the effect of $g_k$ is small in the present model, 
this term gives essentially new contribution to the thermal conductivity, which is beyond the SB relations.
It is to be noted that $g_k$ does not appear when one type of interaction is considered.
It has already been pointed out that such terms beyond the SB relations
do not appear in the Hubbard model~\cite{kontani},
and the same argument can be made in this model
by mapping the two bands to the spin direction.
In other words, considering the two types of interactions is essential
to obtain heat current beyond the SB relations.

Finally, we discuss the relationship between the additional contribution to the 
thermal conductivity and the heat current of excitions. 
As mentioned in the introduction, the thermal transport due to excitons is expected to show the drastic features, 
because excitons do not have electronic charges, 
but have energies contributing to the heat current. 
The heat current due to excitons is not included in the previous studies~\cite{zittartz3,kurihara}, 
because they are based on the SB relations.
%Since excitons do not contribute to the electrical conductivity, 
%$\sigma\left(\epsilon\right)$ in Eq.~(\ref{jonson1}) is solely due to the quasiparticles in EI.
%Correspondingly, the thermal conductivity obtained from Eq.~(\ref{jonson3}) (i.e., SB relations) is also due to the quasiparticles.
However, we find the additional thermal conductivity which does not satisfy the SB relations.
Furthermore, additional thermal conductivity is related to the order parameters of EI.
Therefore, it is suggested that the contribution from the heat current carried by excitons 
is included in additional thermal conductivity.
Such theoretical understanding of the exciton's heat current could provide the basis 
for the experimental determination of the EI state.

\section{Conclusion}
We study the thermal conductivity in EI using a simple quasi one-dimensional two-band model with the two types of Coulomb interactions between these bands.
%we developed the method to identify the contributions to thermal conductivity 
%driven by EI on the basis of a two-band model with two types of interactions. 
First, we obtained the heat current operator microscopically
and clarified that it can not be written in the form of an imaginary-time derivative 
of the electric current operator as Eqs.~(\ref{jcoefficient}) and (\ref{jikanbibun}). 
In other words, by considering two types of interactions, 
additional heat current operator owing to the excitonic phase transition exists, 
which is different from previous studies~\cite{zittartz3,kurihara}.
Then, we studied the linear response theory of $L_{11}$, $L_{21}$, and $L_{22}$ within a mean-field scheme,
and clarified that abovementioned additional heat current operator gives contributions 
in $L_{21}$ and $L_{22}$ which are not expressed in the form of 
Eqs.~(\ref{jonson2}) and (\ref{jonson3}), 
or which are beyond the SB relations. 

%Since the additional contributions are related to the order parameters of EI, 
%we identified them as contribution of excitons to heat current.

\begin{acknowledgments}
We are grateful to Prof. Hidenori Takagi and his group members for fruitful discussions.  
This work is supported by Grants-in-Aid for Scientific Research from the Japan Society for the Promotion of Science (Nos. JP20K03802, JP18H01162, and JP18K03482) 
and JST-Mirai Program Grant Number JPMJMI19A1, Japan.
S.T. was supported by the Japan Society for the Promotion of Science thorough the Program for Leading Graduate Schools (MERIT).
\end{acknowledgments}

\appendix

\section{Derivations of Eqs.~(\ref{denryu}) and (\ref{neturyuuenzansi}) together with the Hamiltonian density}

\begin{figure}[t]
 \begin{minipage}{0.49\hsize}
  \begin{center}
   \includegraphics[width=4.2cm]{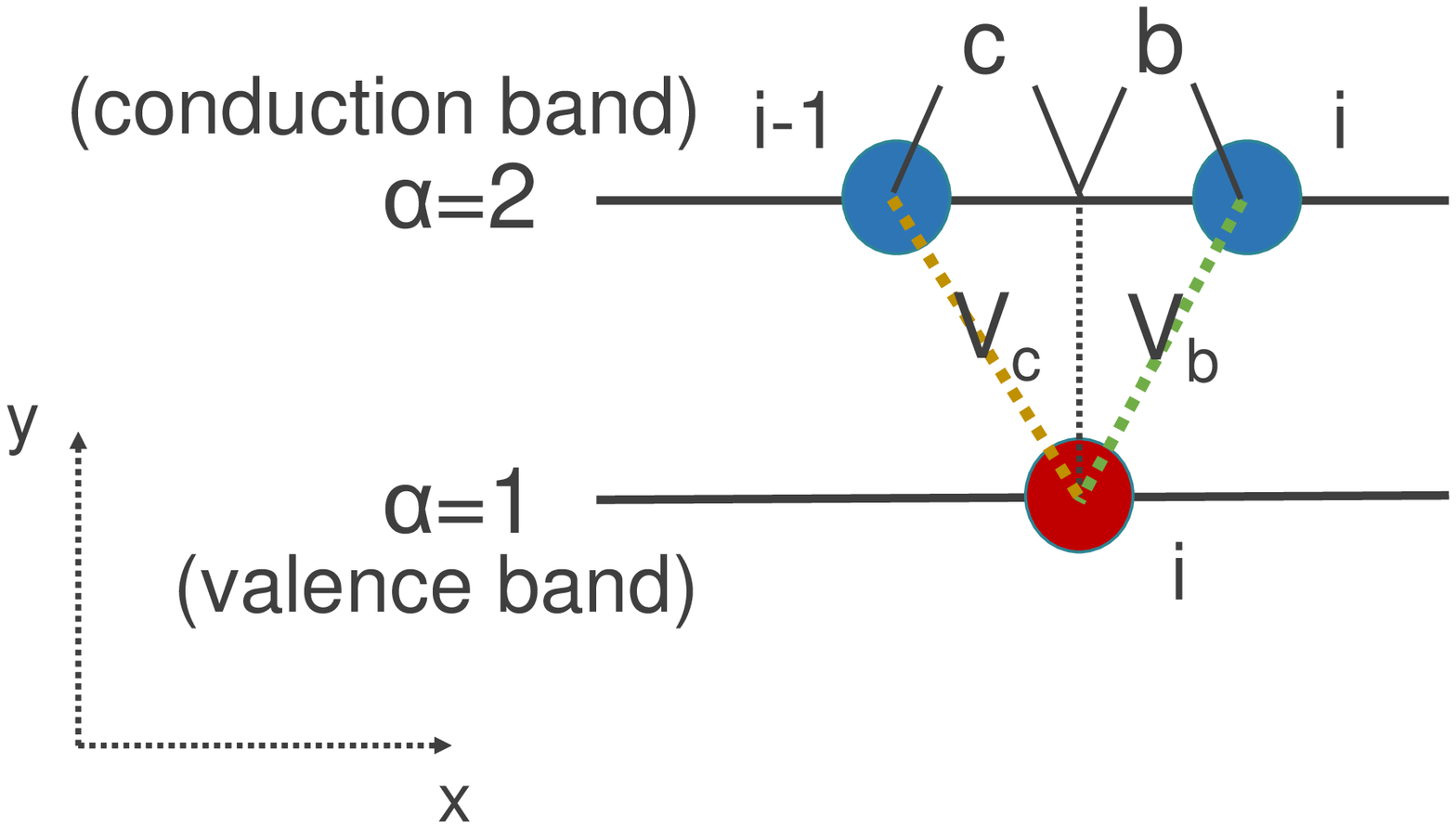}
   \captionsetup{labelformat=empty,labelsep=none}
   \caption*{(a) $\hat{h}_i$}
  \end{center}
 \end{minipage}
 \begin{minipage}{0.49\hsize}
  \begin{center}
   \includegraphics[width=4.2cm]{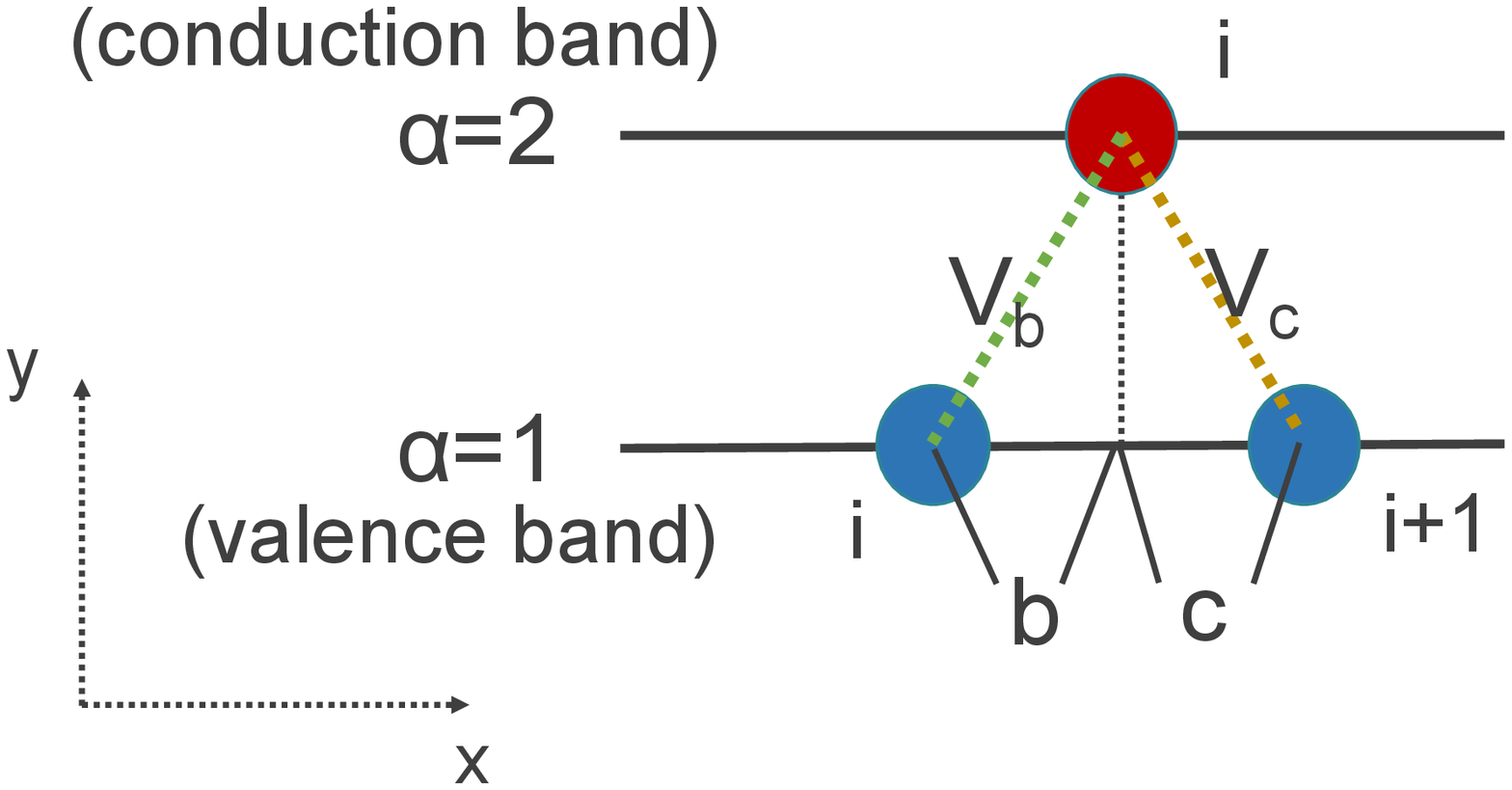}
   \captionsetup{labelformat=empty,labelsep=none}
   \caption*{(b) $\hat{h}_{i+\frac{1}{2}}$}
  \end{center}
 \end{minipage}
\captionsetup{justification=raggedright}
\caption{Schematic picture of the local (a) $i$-th site, (b) ($i$+1/2)-th site.
For the local $i$-th site (($i$+1/2)-th site),
we consider hopping in both directions from the $i$-th site in band 1 (band 2),
and hopping between the ($i$-1)-th site and the $i$-th site in band 2
(between the $i$-th site and the ($i$+1)-th site in band 1).}
\label{fig:local}
\end{figure}

In the preset model (Eq.~(\ref{H})), the Hamiltonian density, $\hat{h}_i$ and $\hat{h}_{i+\frac{1}{2}}$, are obtained as
\begin{align}
\hat{h}_i=& \left(
\frac{t}{2}\hat{c}^{\dagger}_{1,i}\, \hat{c}_{1,i+1}
+\frac{t}{2}\hat{c}^{\dagger}_{1,i-1}\, \hat{c}_{1,i}
-t' \hat{c}^{\dagger}_{2,i-1}\, \hat{c}_{2,i}
+ {\rm h.c.} \right)
\nonumber\\
&+\left( -\epsilon -\mu  \right) \hat{n}_{1,i}
 + \frac{1}{2} \left( \epsilon -\mu \right)
\left( \hat{n}_{2,i} + \hat{n}_{2,i-1} \right)
\nonumber\\
&+ V_b \,\hat{n}_{1,i} \,\hat{n}_{2,i}
+ V_c\, \hat{n}_{1,i} \,\hat{n}_{2,i-1}
\nonumber\\
&+ \sum_{j} \bigl\{V_{{\rm imp}}\left( x_{1,i}-X_j \right)
\hat{n}_{1,i}
+  V_{{\rm imp}}\left( x_{2,i}-X_j \right)\frac{\hat{n}_{2,i}}{2}
\nonumber\\
&+V_{{\rm imp}}\left( x_{2,i-1}-X_j \right)\frac{\hat{n}_{2,i-1}}{2}\bigr\},
\label{localh}
\\
\hat{h}_{i+\frac{1}{2}}=& \left(
t \hat{c}^{\dagger}_{1,i}\, \hat{c}_{1,i+1}
-\frac{t'}{2}\hat{c}^{\dagger}_{2,i-1}\, \hat{c}_{2,i}
-\frac{t'}{2}\hat{c}^{\dagger}_{2,i}\, \hat{c}_{2,i+1}
+ {\rm h.c.} \right)
\nonumber\\
&+\frac{1}{2}\left( -\epsilon -\mu  \right)
\left(\hat{n}_{1,i+1}+\hat{n}_{1,i}\right)
 + \left( \epsilon -\mu \right) \hat{n}_{2,i}
\nonumber\\
&+ V_b \,\hat{n}_{1,i} \,\hat{n}_{2,i}
+ V_c\, \hat{n}_{1,i+1} \,\hat{n}_{2,i}
\nonumber\\
&+ \sum_{j} \bigl\{V_{{\rm imp}}\left( x_{1,i+1}-X_j \right)\frac{\hat{n}_{1,i+1}}{2}
\nonumber\\
&+V_{{\rm imp}}\left( x_{1,i}-X_j \right)\frac{\hat{n}_{1,i}}{2}
+V_{{\rm imp}}\left( x_{2,i}-X_j \right)
\hat{n}_{2,i}\bigr\}.
\label{localh2}
\end{align}
Figure \ref{fig:local} shows the schematic picture of the Hamiltonian density.
Using Eqs.~(\ref{con_eq}), (\ref{localh}) and (\ref{localh2}),
the electric and the heat current density operators are expressed as
\begin{align}
\begin{split}
\frac{ie}{\hbar} \left[ \frac{ \hat{n}_{1,i}+\hat{n}_{2,i} }{a}
,\hat{\mathscr{H}} \right]
=& \frac{2}{a} \left( \hat{j}_{i+\frac{1}{4}}
- \hat{j}_{i-\frac{1}{4}}  \right)
,\\
\frac{ie}{\hbar} \left[ \frac{ \hat{n}_{1,i+1}+\hat{n}_{2,i} }{a}
,\hat{\mathscr{H}} \right]
=& \frac{2}{a} \left( \hat{j}_{i+\frac{3}{4}}
- \hat{j}_{i+\frac{1}{4}}  \right)
,\\
\frac{i}{\hbar} \left[ \frac{\hat{h}_i}{a} , \hat{\mathscr{H}} \right]
=& \frac{2}{a} \left( \hat{j}^Q_{i+\frac{1}{4}}
- \hat{j}^Q_{i-\frac{1}{4}}  \right)
,\\
\frac{i}{\hbar} \left[ \frac{\hat{h}_{i+\frac{1}{2}}}{a} , \hat{\mathscr{H}} \right]
=& \frac{2}{a} \left( \hat{j}^Q_{i+\frac{3}{4}}
- \hat{j}^Q_{i+\frac{1}{4}}  \right)
.\end{split}
\label{jjq}
\end{align}
Then, the total electric current and the
total heat current operators become
\begin{align}
\hat{J}=& \frac{a}{2}\,\sum_i \left(\hat{j}_{i+\frac{1}{4}}+\hat{j}_{i+\frac{3}{4}}\right)
,\;\;
\hat{J}^Q=& \frac{a}{2}\,\sum_i \left(\hat{j}^Q_{i+\frac{1}{4}}+\hat{j}^Q_{i+\frac{3}{4}}\right)
\end{align}
As a result, we obtain Eqs.~(\ref{denryu}) and (\ref{neturyuuenzansi}).

\section{Discussion on the derivation of heat current operator}

The electric and heat current operators should be uniquely determined 
once the Hamiltonian is determined, 
and it should be determined independently of the mean-field approximation. 
Therefore, the current operators should not be calculated using the 
mean field Hamiltonian of Eq.~(\ref{Hmfk}), 
but should be calculated using the original Hamiltonian of Eq.~(\ref{H}) 
before the mean-field approximation is applied to the heat current operator.

In fact, the electric and heat current operators 
calculated using the mean field Hamiltonian (\ref{Hmfk}), 
$\hat{J}'_{{\rm MF}}$ and $\hat{J}'^Q_{{\rm MF}}$, are 
\begin{align}
\hat{J}'_{{\rm MF}}=&
\sum_{k,\alpha,\alpha'} j_{\alpha\alpha'}\left(k\right)
\hat{c}^{\dagger}_{\alpha,k}\,\hat{c}_{\alpha',k},
\end{align}
with
\begin{align}
&j_{11}\left(k\right)\equiv j_1\left(k\right),
\;\;
j_{22}\left(k\right)\equiv j_2\left(k\right),
\nonumber\\
&j_{12}\left(k\right)=\overline{j_{21}\left(k\right)}
\equiv\frac{iea}{2\hbar}\left(\Delta_be^{ikb}-\Delta_ce^{-ikc}\right),
\end{align}
and
\begin{align}
\hat{J}'^Q_{{\rm MF}}=&
\sum_{k,\alpha,\alpha'}j^Q_{{\rm MF}\alpha\alpha'}\left(k\right)
\hat{c}^{\dagger}_{\alpha,k}\,\hat{c}_{\alpha',k}
\nonumber\\
&+\sum_{k,q,\alpha,\alpha'} j^Q_{{\rm imp}\alpha\alpha'}\left(k,q\right)\,
\hat{c}^{\dagger}_{\alpha,k+q}\,\hat{c}_{\alpha',k}
,\end{align}
with 
\begin{align}
&j^Q_{{\rm MF}11}\left(k\right)\equiv
j^Q_{11}\left(k\right)-\frac{a}{\hbar}\Delta_b\,\Delta_c\,{\rm sin}ka,
\nonumber\\
&j^Q_{{\rm MF}22}\left(k\right)\equiv
j^Q_{22}\left(k\right)-\frac{a}{\hbar}\Delta_b\,\Delta_c\,{\rm sin}ka,
\nonumber\\
&j^Q_{{\rm MF}12}\left(k\right)=\overline{j^Q_{{\rm MF}21}}\left(k\right)
\nonumber\\
\equiv &-\frac{a}{\hbar}\left(t-t'\right)\Delta_k\,{\rm sin}ka
\nonumber\\
&+\frac{ia}{2\hbar}\left(\Delta_be^{ikb}-\Delta_ce^{-ikc}\right)
\left\{\left(t-t'\right){\rm cos}ka-\mu\right\},
\nonumber\\
&j^Q_{{\rm imp}11}\left(k\right)\equiv j^Q_{{\rm imp}1}\left(k\right),
\;\;
j^Q_{{\rm imp}22}\left(k,q\right)\equiv j^Q_{{\rm imp}2}\left(k\right),
\nonumber\\
&j^Q_{{\rm imp}12}\left(k,q\right)=\overline{j^Q_{{\rm imp}21}}\left(k\right)
\nonumber\\
\equiv&\frac{1}{2}\Delta_b\,e^{ikb}\left(1+e^{iqb}\right)-\frac{1}{2}\Delta_c\,e^{-ikc}\left(1+e^{-iqc}\right)
.\label{heikinbaneturyu}
\end{align}
These expressions are different from Eq.~(\ref{jcoefficient}). 
Furthermore, if we use these operators $\hat{J_{\rm MF}'}$ and $\hat{J'^{Q}_{\rm MF}}$, 
we obtain various additional terms compared with Eq.~(\ref{l}).

Like this, it is necessary to pay attention to the order of 
the calculation of the heat current operator and the application of the 
mean-field approximation.

\section{Derivations of Eqs.~(\ref{l})}

When we define $\Phi^{\alpha\beta}_{\gamma\delta} \left(k,\omega\right)$ 
as an analytic continuation (
$i\omega_{\lambda}\rightarrow \hbar\omega+i\delta$) 
to 
$\sum_n\mathscr{G}_{\alpha \beta}\left(k\, ,i\epsilon_n\right)
\mathscr{G}_{\gamma \delta}\left(k\, ,i\epsilon_n-i\omega_{\lambda}\right)$
, the product of the Green's function, 
we can calculate Eq.~(\ref{l11keisanhou}) 
by calculating 
\begin{align}
\label{sigmateigi}
\sigma^{\alpha\beta}_{\gamma\delta}\left(k\right) \equiv
\left.\frac{\Phi^{\alpha\beta}_{\gamma\delta}\left(k,\omega\right)
-\Phi^{\alpha\beta}_{\gamma\delta}\left(k,\omega =0\right)}{i\omega}
\right|_{\omega \rightarrow 0}
.\end{align}

Here, we calculate 
$\sigma^{\scriptscriptstyle ++}_{\scriptscriptstyle ++}\left(k\right)$ 
as an example. 
Using the Green's function in Eq.~(\ref{G}) 
and replacing the sum of the Matsubara frequencies 
with the complex integral, we obtain 
\begin{widetext}
\begin{align}
&\sum_n\,\mathscr{G}_{\scriptscriptstyle ++}
\left(k,i\epsilon_n-i\omega_{\lambda}\right)
\mathscr{G}_{\scriptscriptstyle ++}\left(k,i\epsilon_n\right)
\nonumber\\
=&-\frac{1}{2\pi ik_BT}\int^{\infty}_{-\infty}dx
\,\Bigl[\,f\left(x+i\omega_{\lambda}\right)
\Bigl\{ {\scriptstyle 
\frac{x-E_{k-}+i\Gamma_2}
{\left(x-E_{k+}+i\Gamma_1\right)\left(x-E_{k-}+i\Gamma_2\right)+\Gamma_3^2}
\frac{x+i\omega_{\lambda}-E_{k-}+i\Gamma_2}
{\left(x+i\omega_{\lambda}-E_{k+}+i\Gamma_1\right)
\left(x+i\omega_{\lambda}-E_{k-}+i\Gamma_2\right)+\Gamma_3^2}
}\nonumber\\
&{\scriptstyle 
-\frac{x-E_{k-}-i\Gamma_2}
{\left(x-E_{k+}-i\Gamma_1\right)\left(x-E_{k-}-i\Gamma_2\right)+\Gamma_3^2}
\frac{x+i\omega_{\lambda}-E_{k-}+i\Gamma_2}
{\left(x+i\omega_{\lambda}-E_{k+}+i\Gamma_1\right)
\left(x+i\omega_{\lambda}-E_{k-}+i\Gamma_2\right)+\Gamma_3^2}
}\Bigr\} 
\nonumber\\
&+f\left(x\right)
\Bigl\{ {\scriptstyle 
\frac{x-i\omega_{\lambda}-E_{k-}-i\Gamma_2}
{\left(x-i\omega_{\lambda}-E_{k+}-i\Gamma_1\right)
\left(x-i\omega_{\lambda}-E_{k-}-i\Gamma_2\right)+\Gamma_3^2}
\frac{x-E_{k-}+i\Gamma_2}
{\left(x-E_{k+}+i\Gamma_1\right)
\left(x-E_{k-}+i\Gamma_2\right)+\Gamma_3^2}
}\nonumber\\
&{\scriptstyle 
-\frac{x-i\omega_{\lambda}-E_{k-}-i\Gamma_2}
{\left(x-i\omega_{\lambda}-E_{k+}-i\Gamma_1\right)
\left(x-i\omega_{\lambda}-E_{k-}-i\Gamma_2\right)+\Gamma_3^2}
\frac{x-E_{k-}-i\Gamma_2}
{\left(x-E_{k+}-i\Gamma_1\right)
\left(x-E_{k-}-i\Gamma_2\right)+\Gamma_3^2}
}\Bigr\}\,\Bigr]
\nonumber \\
=&\frac{2}{\pi k_BT}\int^{\infty}_{-\infty}dx
\,f\left( x\right)
\Bigl[ {\scriptstyle 
\frac{\left\{\Gamma_1\left(x-E_{k-}\right)^2
+\Gamma_1\Gamma_2^2-\Gamma_2\Gamma_3^2\right\}}
{\left\{\left(x-E_{k+}+i\Gamma_1\right)
\left(x-E_{k-}+i\Gamma_2\right)+\Gamma_3^2\right\}
\left\{\left(x-E_{k+}-i\Gamma_1\right)
\left(x-E_{k-}-i\Gamma_2\right)+\Gamma_3^2\right\}}
}\nonumber\\
&{\scriptstyle \times
\frac{\left(x-E_{k+}\right)
\left\{\left(x-E_{k-}\right)^2-\left(i\omega_{\lambda}+i\Gamma_2\right)^2\right\}
+\Gamma_3^2\left(x-E_{k-}\right)}
{\left\{\left(x+i\omega_{\lambda}-E_{k+}+i\Gamma_1\right)
\left(x+i\omega_{\lambda}-E_{k-}+i\Gamma_2\right)+\Gamma_3^2\right\}
\left\{\left(x-i\omega_{\lambda}-E_{k+}-i\Gamma_1\right)
\left(x-i\omega_{\lambda}-E_{k-}-i\Gamma_2\right)+\Gamma_3^2\right\}}
}\Bigr]
.\end{align}
%where $f\left(E\right)=\frac{1}{e^{\beta E}+1}$ 
%is the Fermi distribution function, 
%and we used the fact that 
%$\frac{1}{e^{\beta (x+i\omega_{\lambda})}+1}=\frac{1}{e^{\beta x}+1}$. 

By performing analytic continuation 
($i\omega_{\lambda}\rightarrow\hbar\omega+i\delta$), 
we obtain 
\begin{align}
\sigma^{\scriptscriptstyle ++}_{\scriptscriptstyle ++}\left(k\right)=
\frac{4\hbar}{\pi k_BT}\int^{\infty}_{-\infty}dx \,f\left(x\right)
\frac{\begin{array}{l}{\scriptstyle \{\, 
\Gamma_1^2\left(x-E_{k+}\right)\left(x-E_{k-}\right)^6
-\Gamma_1\Gamma_2\Gamma_3^2\left(x-E_{k+}\right)^2\left(x-E_{k-}\right)^3
}\\
{\scriptstyle +\left(3\Gamma_1^2\Gamma_2^2
-2\Gamma_1\Gamma_2\Gamma_3^2-\Gamma_1^2\Gamma_3^2\right)
\left(x-E_{k+}\right)\left(x-E_{k-}\right)^4
+\Gamma_1^2\Gamma_3^2\left(x-E_{k-}\right)^5
\}}\end{array}}
{{\scriptstyle \left\{\left(x-E_{k+}+i\Gamma_1\right)
\left(x-E_{k-}+i\Gamma_2\right)+\Gamma_3^2\right\}^3
\left\{\left(x-E_{k+}-i\Gamma_1\right)
\left(x-E_{k-}-i\Gamma_2\right)+\Gamma_3^2\right\}^3}}
.\end{align}
And for the denominator of the integrand,  
\begin{align}
&\left\{\left(x-E_{k+}+i\Gamma_1\right)
\left(x-E_{k-}+i\Gamma_2\right)+\Gamma_3^2\right\}
\left\{\left(x-E_{k+}-i\Gamma_1\right)
\left(x-E_{k-}-i\Gamma_2\right)+\Gamma_3^2\right\}
\nonumber\\
=&\left\{\left(x-E_1\right)^2+\delta_1^2\right\}
\left\{\left(x-E_2\right)^2+\delta_2^2\right\}
\end{align}
is holds, where 
$E_1=E_{k+}-\frac{\Gamma_3^2}{E_{k+}-E_{k-}}
+\mathcal{O}\left(\Gamma^4\right)$, 
$E_2=E_{k-}+\frac{\Gamma_3^2}{E_{k+}-E_{k-}}
+\mathcal{O}\left(\Gamma^4\right)$, 
$\delta_1=\Gamma_1+\mathcal{O}\left(\Gamma^3\right)$, and 
$\delta_2=\Gamma_2+\mathcal{O}\left(\Gamma^3\right)$, 
here $\Gamma_1$, $\Gamma_2$, and $\Gamma_3$ 
are written together as $\Gamma$.
Since we assume that the effect of impurities is small and 
$\Gamma$ is a minute quantity, 
we can use 
$\frac{\delta^{2i-1}}
{\left\{\left(x-a\right)^2+\delta^2\right\}^i}
\sim\frac{\left(2i-3\right)!!}{\left(2i-2\right)!!}
\,\pi\, \delta\left(x-a\right)
$ and 
$\frac{\delta^{2i-1}\left(x-a \right)}
{\left\{\left(x-a\right)^2+\delta^2\right\}^{i+1}}
\sim-\frac{\left(2i-3\right)!!}{\left(2i\right)!!}
\,\pi \,\delta'\left(x\right)$
to obtain 
\begin{align}
\sigma^{\scriptscriptstyle ++}_{\scriptscriptstyle ++}\left(k\right)
=&\frac{4\hbar}{\pi k_BT}\int^{\infty}_{-\infty}dx\, f\left(x\right)
\Bigl[\,{\scriptstyle 
\frac{\mathcal{O}\left(\Gamma^4\right)}
{\left\{\left(x-E_1\right)^2+\delta_1^2\right\}}
\,+\,\frac{\mathcal{O}\left(\Gamma^4\right)}
{\left\{\left(x-E_2\right)^2+\delta_2^2\right\}}
\,+\,\frac{\mathcal{O}\left(\Gamma^4\right)}
{\left\{\left(x-E_1\right)^2+\delta_1^2\right\}^2}
\,+\,\frac{\mathcal{O}\left(\Gamma^6\right)}
{\left\{\left(x-E_2\right)^2+\delta_2^2\right\}^2}
\,+\,\frac{\mathcal{O}\left(\Gamma^6\right)}
{\left\{\left(x-E_1\right)^2+\delta_1^2\right\}^3}
\,+\,\frac{\mathcal{O}\left(\Gamma^6\right)}
{\left\{\left(x-E_2\right)^2+\delta_2^2\right\}^3}
}\nonumber\\
&{\scriptstyle 
\,+\,\frac{\left(x-E_1\right)
\mathcal{O}\left(\Gamma^4\right)}
{\left\{\left(x-E_1\right)^2+\delta_1^2\right\}^2}
\,+\,\frac{\left(x-E_2\right)
\mathcal{O}\left(\Gamma^4\right)}
{\left\{\left(x-E_2\right)^2+\delta_2^2\right\}^2}
\,+\,\frac{\left(x-E_1\right)
\left\{\Gamma_1^2
\,+\,\mathcal{O}\left(\Gamma^4\right)\right\}}
{\left\{\left(x-E_1\right)^2+\delta_1^2\right\}^3}
\,+\,\frac{\left(x-E_2\right)\mathcal{O}\left(\Gamma^6\right)}
{\left\{\left(x-E_2\right)^2+\delta_2^2\right\}^3}
\,+\,\frac{\mathcal{O}\left(\Gamma^6\right)}
{\left\{\left(x-E_1\right)^2+\delta_1^2\right\}
\left\{\left(x-E_2\right)^2+\delta_2^2\right\}}
}\,\Bigr]
\nonumber\\
\sim&
\frac{4\hbar}{k_BT}\int^{\infty}_{-\infty}dx\,f\left( x\right)
\Bigl[\,{\scriptstyle
\frac{\delta\left(x-E_1\right)}{\delta_1}
\mathcal{O}\left(\Gamma^4\right)
\,+\,\frac{\delta\left(x-E_2\right)}{\delta_2}
\mathcal{O}\left(\Gamma^4\right)
\,+\,\frac{\delta\left(x-E_1\right)}{2\delta_1^3}
\mathcal{O}\left(\Gamma^4\right)
\,+\,\frac{\delta\left(x-E_2\right)}{2\delta_2^3}
\mathcal{O}\left(\Gamma^6\right)
\,+\,\frac{3\delta\left(x-E_1\right)}{8\delta_1^5}
\mathcal{O}\left(\Gamma^6\right)
\,+\,\frac{3\delta\left(x-E_2\right)}{8\delta_2^5}
\mathcal{O}\left(\Gamma^6\right)
}\nonumber\\
&{\scriptstyle 
\,-\,\frac{\delta'\left(x-E_1\right)}{2\delta_1}
\mathcal{O}\left(\Gamma^4\right)
\,-\,\frac{\delta'\left(x-E_2\right)}{2\delta_2}
\mathcal{O}\left(\Gamma^4\right)
\,-\,\frac{\delta'\left(x-E_1\right)}{8\delta_1^3}
\left(\Gamma_1^2+\mathcal{O}\left(\Gamma^4\right)\right)
\,-\,\frac{\delta'\left(x-E_2\right)}{8\delta_2^3}
\mathcal{O}\left(\Gamma^6\right)
\,+\,\frac{\delta\left(x-E_1\right)\delta\left(x-E_2\right)}
{\pi \delta_1 \delta_2}
\mathcal{O}\left(\Gamma^6\right)
}\,\Bigr]
\nonumber\\
=&\frac{\hbar}{2k_BT\Gamma_1}f'(E_{k+})+\mathcal{O}\left(\Gamma\right)
\end{align}

In exactly same way, we obtain 
\begin{align}
&\sigma^{\scriptscriptstyle ++}_{\scriptscriptstyle ++}\left(k\right)
= \frac{\hbar}{2k_BT\Gamma_1}f'\left(E_{k+}\right)
+\mathcal{O}\left(\Gamma\right)
,\;\;
\sigma^{\scriptscriptstyle --}_{\scriptscriptstyle --}\left(k\right)
= \frac{\hbar}{2k_BT\Gamma_2}f'\left(E_{k-}\right)
+\mathcal{O}\left(\Gamma\right)
,\nonumber\\
&\sigma^{\scriptscriptstyle +-}_{\scriptscriptstyle +-}\left(k\right)
= \sigma^{\scriptscriptstyle +-}_{\scriptscriptstyle -+}\left(k\right)
= \sigma^{\scriptscriptstyle -+}_{\scriptscriptstyle +-}\left(k\right)
= \sigma^{\scriptscriptstyle -+}_{\scriptscriptstyle -+}\left(k\right)
= \mathcal{O}\left(\Gamma\right)
,\;\;
{\rm Re}\left( \sigma^{\scriptscriptstyle ++}_{\scriptscriptstyle --}
\left(k\right)\right)=
{\rm Re}\left( \sigma^{\scriptscriptstyle --}_{\scriptscriptstyle ++}
\left(k\right)\right)= 
\mathcal{O}\left(\Gamma\right)
,\nonumber\\
&{\rm Re}\left( \sigma^{\scriptscriptstyle ++}_{\scriptscriptstyle +-}
\left(k\right)\right)
={\rm Re}\left( \sigma^{\scriptscriptstyle ++}_{\scriptscriptstyle -+}
\left(k\right)\right)
={\rm Re}\left( \sigma^{\scriptscriptstyle +-}_{\scriptscriptstyle ++}
\left(k\right)\right)
={\rm Re}\left( \sigma^{\scriptscriptstyle -+}_{\scriptscriptstyle ++}
\left(k\right)\right)
= \mathcal{O}\left(\Gamma\right)
,\nonumber\\
&{\rm Re}\left( \sigma^{\scriptscriptstyle --}_{\scriptscriptstyle +-}
\left(k\right)\right)
={\rm Re}\left( \sigma^{\scriptscriptstyle --}_{\scriptscriptstyle -+}
\left(k\right)\right)
={\rm Re}\left( \sigma^{\scriptscriptstyle +-}_{\scriptscriptstyle --}
\left(k\right)\right)
={\rm Re}\left( \sigma^{\scriptscriptstyle -+}_{\scriptscriptstyle --}
\left(k\right)\right)
= \mathcal{O}\left(\Gamma\right)
.\label{sigma}
\end{align}

Eq.~(\ref{l11keisanhou}) can be calculated by using 
Eq.~(\ref{sigma}), and the result is 
\begin{align}
L_{11}=&-\frac{\hbar}{2aN}\sum_k \,\Bigl[\,
\frac{f'\left(E_{k+}\right)}{\Gamma_1}
\left\{\left|u_{11}\right|^2j_1\left(k\right)
+\left|u_{21}\right|^2 j_2\left(k\right)\right\}^2
\nonumber\\
&+\frac{f'\left(E_{k-}\right)}{\Gamma_2}
\left\{\left|u_{12}\right|^2 j_1\left(k\right)
+\left|u_{22}\right|^2 j_2\left(k\right)\right\}^2
\,\Bigr]+\mathcal{O}\left(\Gamma\right)
.\end{align}
Here, we defined $u_{ij}$ 
as the $(i,j)$ component of $\bm{U}$ in Eq.~(\ref{U}).
By calculating  
Eqs.~(\ref{l21keisanhou}) and (\ref{l22keisanhou}) 
in the same way, we obtain  
\begin{align}
L_{21}
=&-\frac{\hbar}{2aN}\sum_k
\,\Bigl[\,\frac{f'\left(E_{k+}\right)}{\Gamma_1}
\left\{ \left|u_{11}\right|^2 j^Q_{11}\left(k\right)
+\left|u_{21}\right|^2 j^Q_{22}\left(k\right)
+2{\rm Re}\left(u_{21}\overline{u}_{11}\,j^Q_{12}\left(k\right)\right)\right\}
\left\{ \left|u_{11}\right|^2 j_1\left(k\right)
+\left|u_{21}\right|^2 j_2\left(k\right) \right\}
\nonumber\\
&+\frac{f'\left(E_{k-}\right)}{\Gamma_2}
\left\{ \left| u_{12}\right|^2 j^Q_{11}\left(k\right)
+\left| u_{22}\right|^2 j^Q_{22}\left(k\right)
+2{\rm Re}\left(u_{22}\overline{u}_{12}\,j^Q_{12}\left(k\right)\right) \right\}
\left\{ \left|u_{12}\right|^2 j_1\left(k\right)
+\left|u_{22}\right|^2j_2\left(k\right) \right\}
\Bigr]+\mathcal{O}\left(\Gamma\right)
,\\
L_{22}
=&-\frac{\hbar}{2aN}\sum_k  \,\Bigl[\,
\frac{f'\left(E_{k+}\right)}{\Gamma_1}
\left\{\left|u_{11}\right|^2 j^Q_{11}\left(k\right)
+\left|u_{21}\right|^2 j^Q_{22}\left(k\right)
+2{\rm Re}\left( u_{21}\overline{u}_{11}\, 
j^Q_{12}\left(k\right) \right) \right\}^2
\nonumber\\
&+\frac{f'\left(E_{k-}\right)}{\Gamma_2}
\left\{ \left|u_{11}\right|^2 j^Q_{11}\left(k\right)
+\left|u_{21}\right|^2 j^Q_{22}\left(k\right)
+2{\rm Re}\left( u_{22}\overline{u}_{12}\, 
j^Q_{12}\left(k\right) \right) \right\}^2
\Bigr]+\mathcal{O}\left(\Gamma\right)
.\end{align}
Using Eqs.~(\ref{U}), (\ref{jcoefficient}), and (\ref{sigmateigi}), 
we can obtain Eq.~(\ref{l}).
\end{widetext}

\end{document}